# Database and deep-learning scalability of anharmonic phonon properties by automated brute-force first-principles calculations

Masato Ohnishi[1,2,*] Tianqi Deng[3,4], Pol Torres[5], Zhihao Xu[6], Terumasa Tadano[7], Haoming Zhang[3,4], Wei Nong[8], Masatoshi Hanai[9], Zeyu Wang[10], Zhiting Tian[11], Ming Hu[12], Xiulin Ruan[13], Ryo Yoshida[2,14,15], Toyotaro Suzumura[9], Lucas Lindsay[16], Alan J. H. McGaughey[17], Tengfei Luo[6,18], Kedar Hippalgaonkar[8,19,20], and Junichiro Shiomi[1,2,10,21,*]

[1] Institute of Engineering Innovation, The University of Tokyo, Tokyo 113-0032, Japan

[2] The Institute of Statistical Mathematics, Research Organization of Information and Systems, Tachikawa, Tokyo 190-8562, Japan

[3] State Key Laboratory of Silicon and Advanced Semiconductor Materials, School of Materials Science and Engineering, Zhejiang University, Hangzhou 310027, China

[4] Key Laboratory of Power Semiconductor Materials and Devices of Zhejiang Province, Institute of Advanced Semiconductors, ZJU-Hangzhou Global Scientific and Technological Innovation Center, Zhejiang University, Hangzhou 311200, China

[5] Eurecat, Technology Centre of Catalonia, Unit of Applied Artificial Intelligence, Cerdanyola del Vallès, 08290, Spain

[6] Department of Aerospace and Mechanical Engineering, University of Notre Dame, Notre Dame, IN 46556, USA

[7] Research Center for Magnetic and Spintronic Materials, National Institute for Materials Science, Tsukuba 305-0047, Japan

[8] School of Materials Science and Engineering, Nanyang Technological University, Singapore 639798, Singapore

[9] Information Technology Center, The University of Tokyo, Tokyo 113-0032, Japan

[10] Department of Mechanical Engineering, The University of Tokyo, Tokyo 113-0032, Japan

[11] Sibley School of Mechanical and Aerospace Engineering, Cornell University, Ithaca, New York 14853, USA

[12] Department of Mechanical Engineering, University of South Carolina, Columbia, SC 29201, USA

[13] School of Mechanical Engineering and Birck Nanotechnology Center, Purdue University, West Lafayette, IN 47907, USA

[14] The Graduate University for Advanced Studies, SOKENDAI, Tachikawa, Tokyo, 190-8562, Japan

[15] Advanced General Intelligence for Science Program (AGIS), RIKEN-TRIP, Wako, Saitama 351-0198, Japan

[16] Materials Science and Technology Division, Oak Ridge National Laboratory, Oak Ridge, TN 37831, USA

[17] Department of Mechanical Engineering, Carnegie Mellon University, Pittsburgh, Pennsylvania 15213, USA

[18] Department of Chemical and Biomolecular Engineering, University of Notre Dame, Notre Dame, IN 46556, USA

[19] Institute of Materials Research and Engineering, Agency for Science Technology and Research, Innovis, Singapore 138634, Singapore

[20] Institute for Functional Intelligent Materials, National University of Singapore, Singapore 117544, Singapore

[21] RIKEN Center for Advanced Intelligence Project, Tokyo 103-0027, Japan

Understanding the anharmonic phonon properties of crystal compounds—such as phonon lifetimes and thermal conductivities—is essential for investigating and optimizing their thermal transport behaviors. These properties also impact optical, electronic, and magnetic characteristics through interactions between phonons and other quasiparticles and fields. In this study, we develop an automated first-principles workflow to calculate anharmonic phonon properties and build a comprehensive database encompassing more than 6,000 inorganic compounds. Utilizing this dataset, we train a graph neural network model to predict thermal conductivity values and spectra from structural parameters, demonstrating a scaling law in which prediction accuracy improves with increasing training data size. High-throughput screening with the model enable the identification of materials exhibiting extreme thermal conductivities—both high and low. The resulting database offers valuable insights into the anharmonic behavior of phonons, thereby accelerating the design and development of advanced functional materials.

## INTRODUCTION

In recent years, the integration of traditional materials science approaches, rooted in fundamental principles, with data-driven methodologies—collectively known as Materials Informatics (MI)—has rapidly advanced, leading to significant breakthroughs in the development of materials for batteries[1–3], catalysts[4], magnetic systems[5], and beyond. For inorganic materials, large-scale computational databases have served as the backbone of MI efforts, including the Materials Project (2013)[6–8] with data on 170,000 materials, OQMD (2013)[9,10] with 1.2 million materials, and AFLOW (2014)[11] with 3.5 million materials. More recently, a series of emerging databases have expanded this landscape, such as a database dedicated to $Fm\bar{3}m$ cubic structures with over 200,000 entries, the Carolina Materials Database (2020)[12,13], DeepMind's GNoME containing 40 million novel crystal structures (2024)[14], and META's OMat24 with 1.1 billion density functional theory (DFT) calculation entries (2024)[15]. However, these databases primarily focus on crystal structures and properties derived from relatively straightforward calculations, such as electronic band structures and band gaps.

In contrast, databases centered on lattice thermal properties, which dominate heat transport in non-metallic materials, remain relatively scarce. Existing simulation-based resources largely provide harmonic phonon properties or lattice thermal conductivity estimates based on approximations—for example, Phonondb[16] offers harmonic properties for ~10,000 materials, and AFLOW employs the quasiharmonic Debye approximation[17]. Experiment-based databases, such as Starrydata[18] and AtomWork[19], compile thermal conductivity and thermoelectric data from the literature. However, these data are significantly influenced by extrinsic factors such as grain size[20,21], carrier density, composition[22,23], impurities[24,25], defects, strain[26–29], and uncertainty of the measurement[30]. Such factors are often undocumented and difficult to control, posing challenges for reliable predictive modeling. Therefore, a first-principles-based database of anharmonic phonon properties is essential for accurately capturing intrinsic thermal behavior, including phonon lifetimes and thermal conductivity, without relying on empirical assumptions.

Complementing these efforts, a team at Microsoft has recently developed an extensive database of anharmonic phonon properties for approximately 246,000 materials[31], using machine learning potentials[32]. While this represents a significant step forward in material research, the available material space for machine learning potential is limited to relatively simple systems due to the focus on high thermal conductivity as the target property—specifically, binary compounds with up to four atoms per primitive cell and ternary compounds composed of group 13–16 elements with up to seven atoms per primitive cell. Additionally, machine learning potentials are trained on data derived from first-principles calculations; therefore, their ability to inherently discover entirely new materials may be limited. Therefore, there remains a need for a first-principles-based database that spans both simple and structurally complex materials.

A first-principles database of anharmonic phonon properties is valuable not only for predicting thermal behaviors of materials but also for understanding a wide range of other material properties. Phonons interact with various particles and excitations—such as electrons[33], magnons[34,35], photons[36,37], plasmons[38], and polaritons[39,40]—affecting[39,40] mechanical, electrical, electronic, optical, and magnetic properties. This highlights the importance of detailed phonon-property datasets that comprehensively capture vibrational properties of solids, particularly describing anharmonic phonon properties based on theoretical calculations using consistent computational approaches/parameters. Such a database will offer critical insights into diverse material behaviors and accelerate the discovery of novel functional material design.

First-principles approaches for calculating anharmonic phonon properties in condensed materials has been actively pursued for many years, triggered by development of computational methods using DFT around 2010[41–43]. In standard first-principles phonon analysis, three-phonon scattering rates are evaluated via quantum perturbative theories under the relaxation time approximation[44–47] to solve the Boltzmann transport equation (BTE)[48]. This approach has been widely applied and has become a rigorous and foundational numerical application for understanding and predicting thermal transport in materials. Building on this framework, a variety of methods have been developed or integrated into computational packages to enhance the accuracy of phonon property calculations, particularly for systems with extreme thermal transport behaviors. Iterative[46,49,50] and direct[51,52] solutions to the BTE offer improved treatment of phonon-phonon interactions by considering the effects of both normal and Umklapp scattering rates, whereas the relaxation time approximation considers only Umklapp scattering as resistive. Furthermore, four-phonon interactions[53,54] in non-metallic systems have been shown to play a significant role in determining their thermal transport behaviors.

At finite temperatures, phonon renormalization modifies harmonic force constants, a process that can be accounted for using first-order self-consistent phonon theory[55–57] and its improved variant incorporating the bubble self-energy corrections[58]. The phonon gas model, which treats phonons as heat-carrying particles that scatter and propagate like molecules in a gas, is extended by the unified phonon theory—also known as the Wigner heat transport formulation[59]—provides a framework for analyzing phonon transport in both the particle (Peierls transport) and wave (coherent transport) pictures.

In addition to phonon-phonon interactions, other scattering mechanisms and intrinsic factors can also play a significant role in thermal transport. Electron-phonon interactions can be accurately analyzed using first-principles methods[60–62]. Weak and strong impurity scatterings can be effectively treated using the perturbative[24] or T-matrix approaches[63,64], respectively. Additionally, intrinsic structural fluctuations at finite temperatures, particularly in complex compounds, can be captured through a combination of cluster expansion and Monte Carlo simulations[22,23,65]. Although this current study employs a fundamental approach based on three-phonon interactions within the relaxation time approximation, the resulting data provide a solid foundation for advanced calculations including more complex scattering effects.

With advancement of computational methods, the development of thermofunctional materials has been accelerated through the integration of informatics and data science. Early studies in this field employed high-throughput calculations with simplified models to identify materials with Peierls lattice thermal conductivities $\kappa_{\mathrm{p}} \approx 1.0\ [\mathrm{Wm^{-1}K^{-1}}]$ [66], and Bayesian optimization techniques were used to discover materials with $\kappa_{\mathrm{p}} < 0.5\ [\mathrm{Wm^{-1}K^{-1}}]$[67]. However, access to anharmonic phonon property data remains limited. To circumvent this, researchers have used harmonic phonon properties[68] and other material descriptors[69], focusing on specific materials such as half-Heuslers[70] and chalcogenides[71], and have developed thermal conductivity databases based on approximations[72], including the Callaway model[73] and minimum thermal conductivity model[74].

In parallel, various techniques have emerged to estimate higher-order force constants at a practical computational cost as the number of displacement patterns required by the finite-displacement method increases rapidly with the order of the force constants. Approaches such as compressive sensing[56,75], projector-based methods for constructing orthonormal basis sets[76], and machine learning potentials[77,78] have been explored. Furthermore, fine-tuned models[79] derived from foundation models[80] have demonstrated improved accuracy. In addition to force constants, the analysis of high-order anharmonic phonon properties—such as four-phonon scattering and phonon renormalization—remains computationally intensive[81]. To address this, machine learning approaches have been introduced, including transfer learning to estimate four-phonon scattering rates using three-phonon scattering data[82].

Driven by this need for a first-principles-based anharmonic phonon property database and building on recent advancements in phonon analysis, we developed an automated computational framework for first-principles phonon calculations that streamlines the workflow and reduces computational complexity. Using this framework, we constructed a large-scale database comprising anharmonic phonon properties for over 6,000 materials, systematically capturing phonon transport characteristics across a wide range of material classes. Leveraging this dataset, we applied machine learning techniques to predict key anharmonic phonon properties, including Peierls lattice thermal conductivity and its spectral distribution. This integrated approach not only deepens our understanding of anharmonic phonon behavior but also accelerates the data-driven discovery of novel functional materials across various application domains.

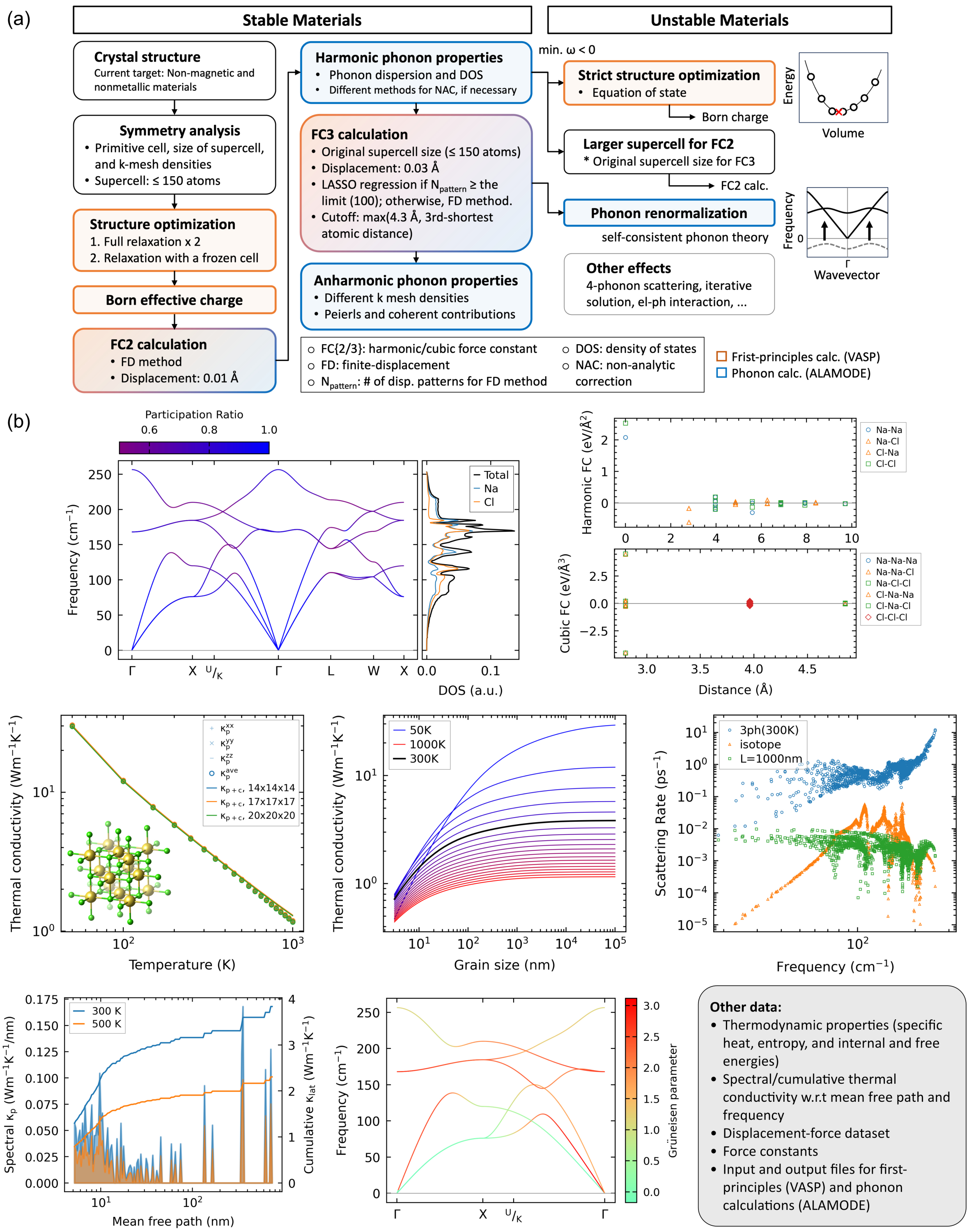

Fig. 1: Automation of anharmonic phonon property calculations using a first-principles approach. (a) Automated workflow implemented in the developed software, auto-kappa. (b) Example output generated by auto-kappa for rock salt NaCl (mp-22862). The results include phonon dispersion with participation ratio and DOS, representative atomic distances for force constants (FCs), temperature- and grain-size-dependent thermal conductivity, mode-dependent phonon scattering rates and lifetimes, spectral and cumulative thermal conductivity as functions of mean free path and frequency, and Grüneisen parameters. In addition, a computational time chart, thermodynamic properties, and various text files—such as displacement–force datasets, force constants, and input/output scripts for simulations—are generated.

## RESULTS AND DISCUSSION

### Automation of Anharmonic Phonon Analysis

We developed automation software named “auto-kappa” (https://github.com/masato1122/auto-kappa) for performing first-principles anharmonic phonon calculations. Given the complexity of phonon analysis, the software automatically addresses key challenges, including precise structure optimization to minimize residual stress and procedures to eliminate imaginary frequencies associated with unstable phonon modes. Specifically, these include structure optimization using an equation of state and increasing the supercell size for force calculations. The automated workflow for anharmonic phonon calculations is summarized in Fig. 1(a), with detailed computational procedures described in the Methods section. While remarkable efforts have been made to automate similar processes for analyzing anharmonic phonon properties[83–85], several challenges were still encountered in the high-throughput calculations of this study. These included the need for automatic adjustment of VASP and ALAMODE parameters (e.g., the cutoff length for force constants and the treatment of the non-analytical correction), job parameters (e.g., the number of parallel processes and the type of parallelization), and the complexity of obtaining relaxed structures, as illustrated in Fig. 1(a). To overcome these challenges, we enhanced auto-kappa to resolve them automatically. Using the developed software, we have calculated the Peierls lattice thermal conductivity ($\kappa_{\mathrm{p}}$) based on the relaxation time approximation as well as the coherence lattice thermal conductivity ($\kappa_{\mathrm{c}}$). Although the software includes an implementation of the self-consistent phonon approach to account for phonon renormalization, the dataset used in this study was generated using the conventional method based on three-phonon interactions within the relaxation time approximation.

Using the developed software, we constructed *Phonix* a database of anharmonic phonon interactions, comprising 7,038 materials. The name *Phonix* highlights its broader scope, extending beyond the phonon–phonon interactions examined here to future coverage of interactions with diverse quasiparticles and nanostructures. The database includes input files, intermediate data, output results, and generated figures—as illustrated in Fig. 1(b). These comprise: phonon dispersion with participation ratio, density of states (DOS), the relationship between force constants and their representative atomic distance (maximum distance among corresponding atoms for the anharmonic case), temperature- and grain-size-dependent thermal conductivity, mode-dependent phonon scattering rates and lifetimes, cumulative and spectral thermal conductivity as a function of mean free path and frequency, Grüneisen parameters, thermodynamic properties (temperature-dependent specific heat, entropy, and internal and free energies), harmonic and anharmonic force constants, displacement-force datasets used for calculating the force constants, and input/output scripts for first-principles (VASP[86]) and phonon (ALAMODE[45]) calculations. Naturally, for materials exhibiting imaginary frequencies, only harmonic property data are included. The target materials in this study consist of all entries from the Phonondb dataset (version 2018-04-07), comprising 10,034 materials, and non-metallic, non-magnetic materials from the Materials Project (version 2022.10.28), comprising 11,418 materials after excluding overlaps with Phonondb. In total, the dataset includes 21,452 unique materials. Although the full phonon analysis has not yet been completed for every material—primarily due to the high computational cost associated with rigorous structural optimization and the use of larger supercells (see Methods for details)—we have successfully calculated anharmonic phonon properties for over 7,000 materials. While we have also obtained a significantly larger set of harmonic phonon properties, including those for materials with unstable phonon modes with imaginary frequencies, this study focuses exclusively on the anharmonic phonon properties, which represent the more compelling aspect of our database. The complete database will be made available on ARIM-mdx[87]. We would also like to emphasize that the database released with this paper represents only the first version, and we are continuously working to improve both the quality and quantity of the data.

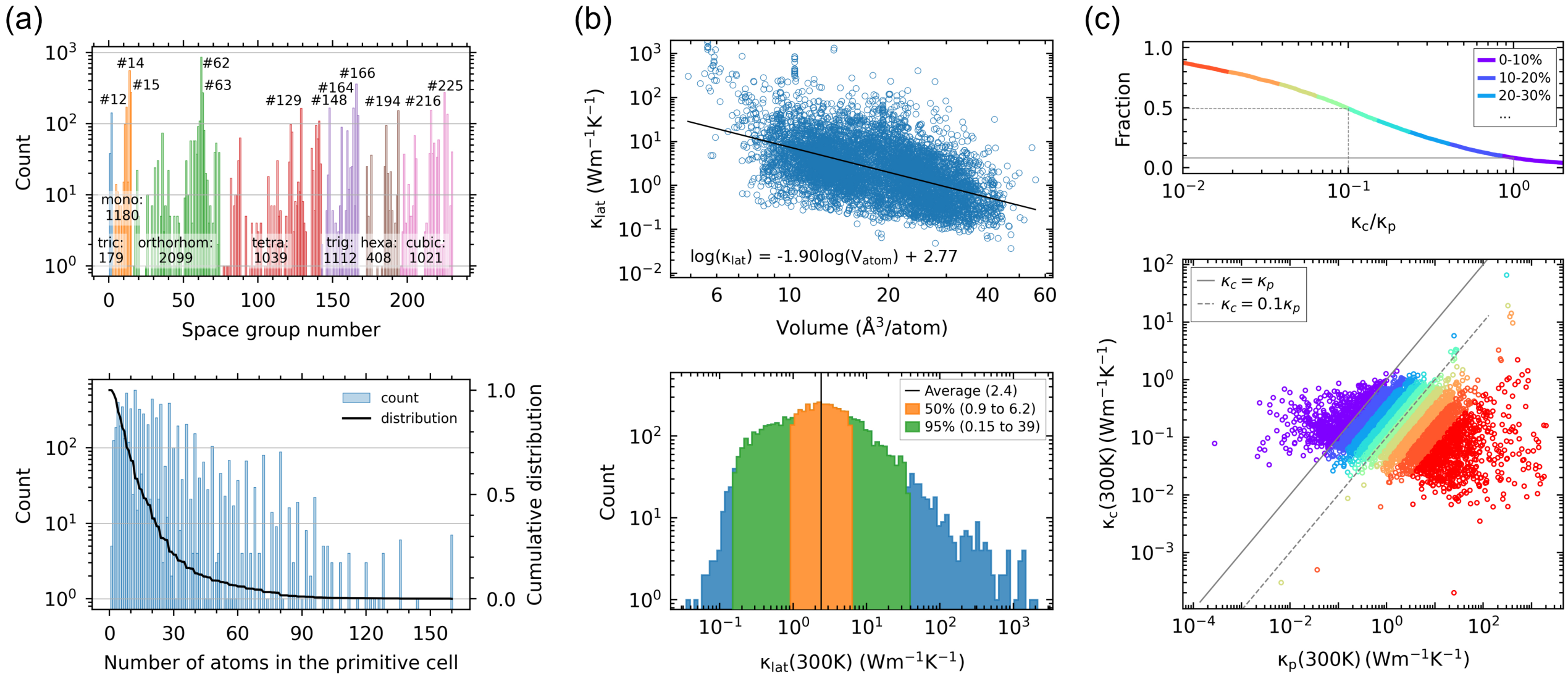

Fig. 2: Data analysis of the *Phonix* database, a database for anharmonic phonon interaction, comprising 7,038 materials. (a) Distribution of space groups and crystal systems (top), and the number of atoms in the primitive cell (bottom) for the crystal structures in the database. (b) Relationship between lattice thermal conductivity ($\kappa_{\text{lat}}$) and volume per atom, along with the distribution of $\kappa_{\text{lat}}$ at 300 K. (c) Comparison of the Peierls ($\kappa_{\text{p}}$) and coherence ($\kappa_{\text{c}}$) contributions to $\kappa_{\text{lat}}$. Solid and dotted lines represent $\kappa_c = \kappa_p$ and $\kappa_c = 0.1 \times \kappa_p$, respectively.

## Database Analysis

First, we analyzed the crystal structures of the materials for which anharmonic phonon properties were computed. As shown in Fig. 2(a), the dataset encompasses a wide range of materials. Among the most populated space groups, as shown at the top of Fig. 2(a), space group 14 includes quartz-like structures such as $SiO_2$; space group 62 includes the anatase phase of $TiO_2$, commonly used as a photocatalyst; space group 166 contains well-known topological insulators and thermoelectric materials like $Bi_2Te_3$ and $Bi_2Se_3$; and space group 225 comprises rock salt structures such as NaCl, PbTe, and PbSe.

Although the current dataset is limited to non-metallic and non-magnetic materials, it is not constrained by the size of the primitive cell, as shown in the bottom panel of Fig. 2(a). Some materials include more than 100 atoms, with the maximum reaching 160 atoms. Among these, five out of seven materials with the highest atom counts belong to space group 62. However, most materials in the database contain fewer than 30 atoms, with half containing fewer than 16 atoms.

Regarding elemental diversity, the Phonix materials contain elements from a broad range of groups, as shown in Supplementary Fig. S1. While transition metals appear less frequently—likely due to the exclusion of magnetic materials in the current version of Phonix—and group 18 elements are present only as single-element systems, all elements from periods 1 to 6 and groups 1 to 17 of the periodic table, except for Po and At, are represented in the Phonix materials. The broad diversity in space groups and structural complexity highlights the versatility of the database as a platform for exploring and developing a wide spectrum of inorganic materials. Notably, only 283 out of 7,038 crystal structures (4.0%) satisfy the search criteria employed by the Microsoft database (MatterK[31]) , while the specific materials contained in their database are not publicly accessible. We believe that both types of databases play complementary and essential roles: databases based on first-principles calculations are crucial for expanding our knowledge toward unexplored materials, while those based on machine learning potentials are important for interpolating data within the known materials space.

Subsequently, the distribution of thermal conductivity was analyzed. Throughout this paper, we used the thermal conductivity at 300 K obtained using the densest **q**-mesh in auto-kappa—1500 q-points·$\mathrm{Å}^3/\mathrm{atom}$—for all discussions. Lattice thermal conductivity ($\kappa_{\mathrm{lat}}$) generally decreases with increasing volume per atom ($V_{\mathrm{atom}}$)[67]. According to Phonix, $\kappa_{\mathrm{lat}}$, including both the Peierls ($\kappa_{\mathrm{p}}$) and coherence ($\kappa_{\mathrm{c}}$) contributions, at 300 K exhibited the following relationship: $\log_{10}(\kappa_{\mathrm{lat}}) \propto \alpha \log_{10}(V_{\mathrm{atom}})$ , where the coefficient $\alpha$ was estimated to be $-1.91$, as illustrated in Fig. 2(b). The average $\kappa_{\mathrm{lat}}$ at 300 K was 2.4 $\mathrm{Wm^{-1}K^{-1}}$, as shown in Fig. 2(b). Half of the materials exhibited $\kappa_{\mathrm{lat}}$(300 K) values between 0.95 and 6.2 $\mathrm{Wm^{-1}K^{-1}}$, while 95% fell within the range of 0.15 to 39 $\mathrm{Wm^{-1}K^{-1}}$. Among the high-thermal-conductivity (high-$\kappa$) materials, 0.14% (10 materials) exhibited $\kappa_{\mathrm{lat}} >$ 1,000 $\mathrm{Wm^{-1}K^{-1}}$, 0.35% (24) exceeded 500 $\mathrm{Wm^{-1}K^{-1}}$, and 0.92% (65) exceeded 200 $\mathrm{Wm^{-1}K^{-1}}$, as listed in Table S1. In the list of calculated materials exhibiting $\kappa_{\mathrm{lat}} > 200\ \mathrm{Wm^{-1}K^{-1}}$, shown in Supplementary Fig. S2, the majority (28 out of 65) were polymorphs of carbon or SiC. Meanwhile, to the best of our knowledge, the following materials, including their polymorphs, have not been synthesized experimentally and have rarely been discussed as high-$\kappa$ materials: triclinic $Hg(BiS_2)_2$ (ID: mp-554921, space group: 12, $\kappa_{p,\{xx,yy,zz\}}$ = 292, 2.5, and 943 $\mathrm{Wm^{-1}K^{-1}}$), cubic HC (mp-1079612, 199, $\kappa_{p,ave} = 306\ \mathrm{Wm^{-1}K^{-1}}$), cubic BiB (mp-1006880, 216, $\kappa_{p,ave} = 235\ \mathrm{Wm^{-1}K^{-1}}$), and trigonal $CsHoS_2$ (mp-505158, 166, $\kappa_{p,\{zz,xx(yy)\}}$ = 22 and 657 $\mathrm{Wm^{-1}K^{-1}}$). It is also noteworthy that the triclinic $Hg(BiS_2)_2$ and trigonal $CsHoS_2$ exhibits highly anisotropic heat conduction with $\kappa_{p,zz}/\kappa_{p,yy} = 392$ and $\kappa_{p,xx(yy)}/\kappa_{p,zz} = 30$, respectively. The value of $Hg(BiS_2)_2$ is comparable to—or even exceeds—that of graphite[88]. On the other hand, among low-$\kappa$ materials, 0.21% (14 materials) exhibited $\kappa_{\mathrm{lat}} < 0.1\ \mathrm{Wm^{-1}K^{-1}}$ (see Supplementary Fig. S3), 15% (1036) exhibited $\kappa_{\mathrm{lat}} < 0.5\ \mathrm{Wm^{-1}K^{-1}}$, 28% (1912) exhibited $\kappa_{\mathrm{lat}} < 1.0\ \mathrm{Wm^{-1}K^{-1}}$, and 72% (4965) exhibited $\kappa_{\mathrm{lat}} < 5.0\ \mathrm{Wm^{-1}K^{-1}}$. Considering that finding materials with $\kappa_{\mathrm{p}} \approx 0.5\ \mathrm{Wm^{-1}K^{-1}}$ was challenging in pioneering studies[67], the obtained dataset provides significant amount of information on low-$\kappa$ materials. While phonon renormalization and four-phonon scattering should be considered for accurately calculating small $\kappa_{\mathrm{lat}}$, this analysis suggests that identifying low-$\kappa$ materials may be relatively easier than finding high-$\kappa$ materials, which remains a greater challenge.

Moreover, it is insightful to compare the Peierls and coherent contributions to the total lattice thermal conductivity. In most materials, particularly high-$\kappa$ materials, the coherent contribution is smaller than the Peierls contribution or sometimes even negligible. However, we observed that a considerable number of materials exhibited a significant coherent contribution: $\kappa_c \geq \kappa_p$ in 8.1% of materials (purple regions in the top and bottom panels of Fig. 2(c), bounded by solid lines), and $\kappa_c \geq 0.1 \times \kappa_p$ in 49%, nearly half of the dataset (bluish regions, bounded by dotted lines). While the relative contribution of the coherent component is known to have a significant effect when the Peierls contribution is small, a large $\kappa_c$ was obtained for SiC polymorphs, which are located in the top-right corner ($\kappa_p \approx 200$ to 500 and $\kappa_c > 10\ \mathrm{Wm^{-1}K^{-1}}$) of the bottom panel of Fig. 2(c). Although the relative contribution of $\kappa_c$ remains small compared to the Peierls conductivity, it is interesting that high-$\kappa$ materials, SiC[89–91], may exhibit a large coherent phonon conductivity ($> 10\ \mathrm{Wm^{-1}K^{-1}}$ and up to 60 $\mathrm{Wm^{-1}K^{-1}}$ at 300 K). Since SiC has more than 200 polymorphs[92], and some of them contain a substantial number of atoms ($> 50$), the densely packed phonon branches resulting from the large number of atoms lead to a large $\kappa_c$, as shown in Supplementary Fig. S4. The developed database contains 15 polymorphs of SiC, among which $Si_{36}C_{36}$ exhibits the highest $\kappa_c$ of 65 $\mathrm{Wm^{-1}K^{-1}}$, while its $\kappa_p$ reaches $\kappa_{p,xx(yy)} = 305$ and $\kappa_{p,zz} = 11\ \mathrm{Wm^{-1}K^{-1}}$.

### Computational Accuracy

To assess computational validity, we compared the results obtained in this study with those in Phonondb[93] as well as with experimental thermal conductivity data. As shown in Supplementary Fig. S5, the phonon dispersions calculated in this work exhibit excellent agreement with those reported in Phonondb. The remaining discrepancies are likely attributable to differences in the relaxed structures, particularly the lattice constants. Overall, this comparison further supports the reliability of both datasets. In addition to harmonic properties, an anharmonic phonon property—namely, the lattice thermal conductivity at room temperature—was compared with experimental data for 103 single-crystal compounds. While calculated data deviate from experimental data for certain materials, calculated data overall show good agreement with the experimental results, as shown in Fig. S6. To further reduce the discrepancies between computational and experimental values, additional factors should be considered in the simulations, including spin–orbit interaction, long-range interactions[94–96], four-phonon scattering, and others. It is also worth noting that these discrepancies could potentially be reduced by employing machine-learned surrogate models, particularly in cases with substantial deviations, as informed by our experience.

Computational accuracy may be limited for materials with high lattice thermal conductivity and should be interpreted with caution, as the study prioritized generating a large dataset under constrained computational resources. The automated calculations occasionally produce excessively high thermal conductivity values—exceeding several thousand $\mathrm{Wm^{-1}K^{-1}}$—which apparently to be unrealistic at this point. These overestimations typically arise from flat phonon bands or acoustic branches. In some instances, phonon modes on flat optical branches exhibit abnormally long lifetimes, while in others, low-frequency acoustic modes display either excessively long lifetimes or unusually high group velocities, as illustrated in Supplementary Fig. S7. To achieve more accurate thermal conductivity estimates, larger supercell sizes (up to 200 atoms) and/or denser **q**-point meshes are required. Another crucial factor is the inclusion of four-phonon interactions, which are expected to reduce the overestimated phonon lifetimes. Although the direct calculation of four-phonon scattering rates is computationally demanding, employing machine learning techniques to predict their effects[82] represents a promising future direction for enhancing the database. In the subsequent machine learning analysis of anharmonic phonon properties, such implausible data have been excluded. Details regarding the computational accuracy of first-principles phonon analysis—including the effects of supercell size and the methods used to obtain force constants—are provided in Supplementary Sec. VIII.

### Deep Learning Scaling Law for Anharmonic Phonon Properties

Using the database developed in this study, we conducted machine learning predictions for anharmonic phonon properties and investigated how prediction accuracy scales with data size[14,97–100]. Our database enables the machine learning prediction of spectral thermal conductivity, not merely scalar values such as $\kappa_{\mathrm{lat}}$ at room temperature (300 K). Since modal lattice thermal conductivity depends on mean free path (MFP) and phonon frequency, predicting spectral thermal conductivity is essential for evaluating the effects of nanostructuring[101,102] and interactions with other particles and excitations, including electrons[33], photons[36,37], and magnons[34,35]. Here, we demonstrate predictions for Peierls thermal conductivity ($\kappa_{\mathrm{p}}$) and cumulative Peierls thermal conductivity ($\kappa_{\mathrm{cumul}}$) as functions of MFP ($\Lambda$) at 300 K. Additional examples of spectral thermal conductivity predictions as functions of frequency and the maximum phonon frequency are provided in the Supplementary Information (see Supplementary Fig. S10 and related discussion).

In this study, we employed the crystal graph convolutional neural network (CGCNN)[103] to predict scalar quantities, such as thermal conductivity, and the Euclidean neural network (e3nn)[104] to predict spectral functions. In CGCNN, atoms are represented by node features composed of one-hot encodings of nine atomic properties, including group number, period number, and electronegativity, while interatomic distances are encoded as discretized edge features. In e3nn, atomic species are represented by 118-dimensional mass-weighted one-hot vectors, and interatomic relations are described using relative position vectors. The e3nn framework incorporates the SE(3)-Transformer[105]—a state-of-the-art architecture for three-dimensional point clouds and graphs—which is equivariant under continuous 3D roto-translations and rigorously accounts for structural symmetries, including mirror (O(3)) and rotational (SO(3)) symmetries, both of which are crucial for phonon analysis. This method has recently been applied to the prediction of complex phonon properties, including DOS[106] and phonon dispersion[107,108]. Further methodological details are provided in the Methods section.

By performing machine learning predictions for $\kappa_{\mathrm{p}}$ and normalized $\kappa_{\mathrm{cumul}}$ ($\kappa_{\mathrm{cumul}}^{\mathrm{norm}}(\Lambda)$) at 300 K using various training dataset sizes ($N_{\mathrm{train}}$), we observed clear scaling behavior with respect to data size, as shown in the left panels of Figs. 3(a) and 3(b). These results clearly demonstrate the enhancement in prediction accuracy enabled by our database. The relationship between mean absolute error (MAE) and $N_{\mathrm{train}}$ was fitted using the empirical formula[98]: $(\mathrm{error}) = (N_c/N_{\mathrm{train}})^{\alpha}$ $(N_c, \alpha > 0)$, where $N_C$ is a constant and $\alpha$ is the scaling factor indicating how effectively increased data improves predictive accuracy. The scaling factors were 0.17 for $\kappa_p$ and 0.14 for $\kappa_{\mathrm{cumul}}$, as shown in Figs. 3(a) and 3(b), and ranged from 0.077 to 0.30 for other properties, as illustrated in Supplementary Fig. S6. These values are comparable to those for large language models (0.095)[98] and force prediction tasks in crystalline materials (0.21)[14] (see Supplementary Fig. S6(e)). As the database continues to expand, the predictive accuracy of surrogate models for large-scale materials screening is expected to improve further. For example, according to the fitted scaling law, the MAE for $\log_{10}\kappa_{\mathrm{p}}$ is expected to decrease to 0.15 as the training dataset size approaches $2.3 \times 10^5$. Nevertheless, brute-force calculations of anharmonic phonon properties for $10^5$-order materials—particularly including higher-order effects such as four-phonon scattering and phonon renormalization—remain impractical. Therefore, further expansion of the database will require machine learning-based acceleration methods, such as machine learning potentials[31,77], to facilitate the efficient evaluation of phonon properties[76,82,109].

The right panels in Figs. 3(a) and 3(b) show representative test cases selected from 50 ensembles for each data size, chosen as those with MAE values closest to the average for the corresponding condition. For instance, when $N_{\mathrm{all}} = 1{,}000$, where $N_{\mathrm{all}}$ denotes the total number of data points used for training, validation, and testing, the average MAE of $\log_{10}\kappa_{\mathrm{p}}$ was 0.37, as shown in the left panel of Fig. 3(a). The middle panel on the right side of Fig. 3(a) displays a representative case with an MAE of 0.377. The prediction results in the right panel of Fig. 3(a) clearly demonstrate that the predicted data points cluster more closely around the parity line as $N_{\mathrm{all}}$ increases. Similarly, the right panel of Fig. 3(b) shows that the fluctuations in the predicted curve are reduced with increasing $N_{\mathrm{all}}$, and the predicted trend aligns more closely with the first-principles results (grey line) for larger datasets.

The exceptional predictive performance for $\kappa_{\mathrm{cumul}}$ is emphasized in Fig. 3(c). As shown in the left panel, 50% (75%) of the test data yielded an MAE for $\log_{10}\kappa_{\mathrm{cumul}}^{\mathrm{norm}}$ below 0.04 (0.09). This panel illustrates the MAE distribution, while the right panels provide prediction examples for individual materials. In the right panel, 50% of the predicted curves exhibit excellent agreement (green and blue regions) with the first-principles results (black line), while 75% demonstrate good agreement (orange region). Even for the final group, where MAE exceeds 0.09 (red region), although the initial value of $\kappa_{\mathrm{cumul}}$—i.e., the $\kappa_{\mathrm{p}}$ contribution from phonons with MFPs shorter than 1 nm—shows a discrepancy, the MFP range where $\kappa_{\mathrm{cumul}}$ begins to increase remains reasonably well predicted.

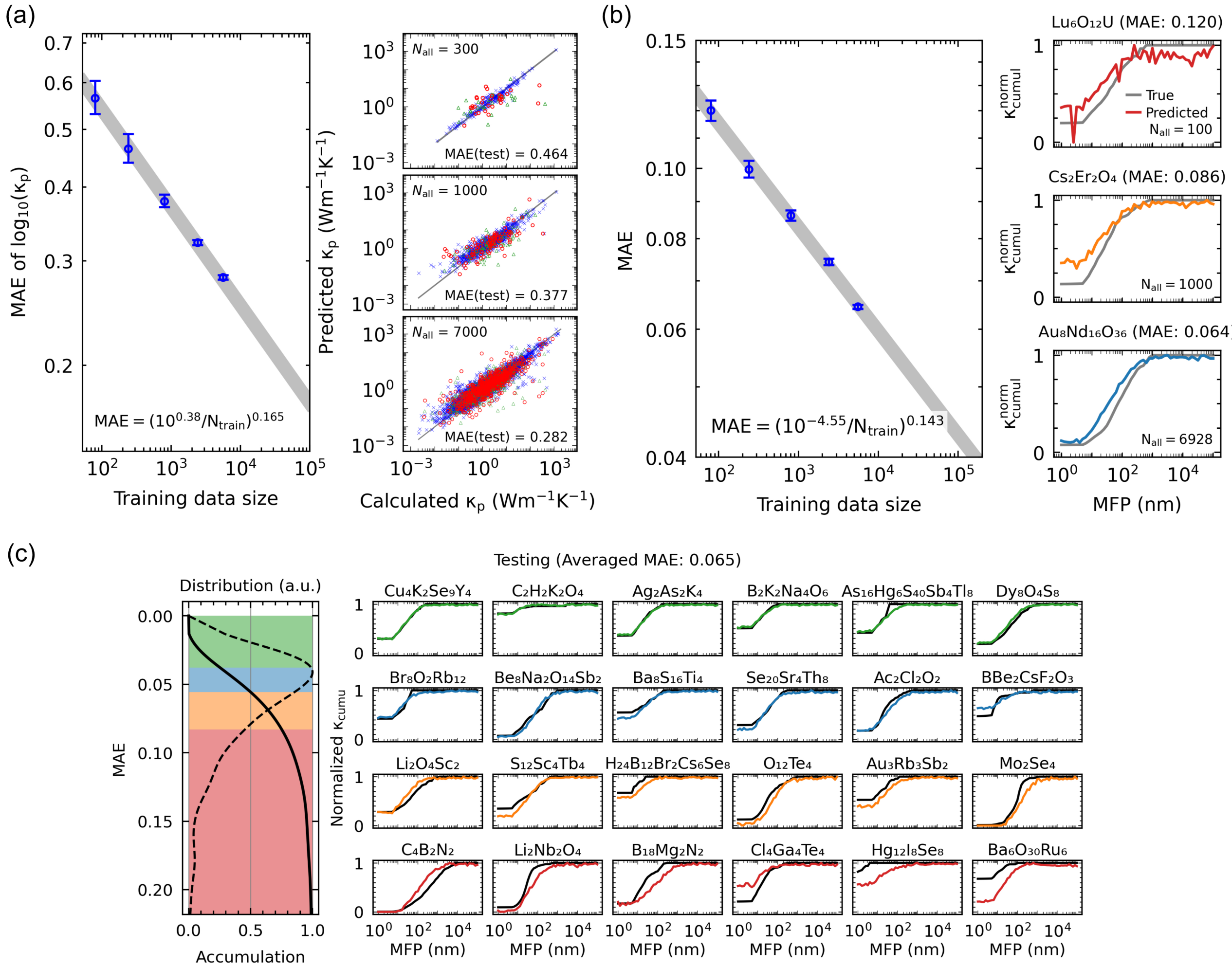

Fig. 3: Deep learning scaling law for anharmonic phonon properties as a function of training data size. (a) Peierls thermal conductivity ($\kappa_\mathrm{p}$) and (b, c) normalized cumulative Peierls thermal conductivity ($\kappa_\mathrm{cumul}^\mathrm{norm}$) were predicted using graph neural networks. The left panels in (a) and (b) show the reduction of mean absolute error (MAE) with increasing data size, demonstrating clear scaling behavior. MAEs were evaluated using $\log_{10}\kappa_\mathrm{p}$ and $\kappa_\mathrm{cumul}^\mathrm{norm}$, respectively. The fitted scaling curve is shown as a grey line, with the corresponding equation displayed at the bottom of each panel. Error bars represent the 90% confidence interval based on 50 ensembles. The right panels in (a) and (b) show prediction examples at different data sizes, selected based on MAE values closest to the ensemble average. In panel (a), blue, green, and red markers represent training, validation, and test data, respectively. In panel (b), colored lines indicate predicted results, while grey lines show data from first-principles calculations (c) Prediction results for $\kappa_\mathrm{cumul}^\mathrm{norm}$ using the entire dataset ($N_\mathrm{all} \approx 5{,}000$). The left panel presents the MAE distribution (dotted line) and its cumulative sum (solid line), color-coded by quartile. The right panels display multiple examples of predicted $\kappa_\mathrm{cumul}^\mathrm{norm}$ curves; colored lines indicate predictions, and black lines represent reference calculations.

## Screening using the Phonix database

Using prediction models developed from our database, we screened materials with high and low thermal conductivity from the GNoME database[14], which contains 381,000 novel crystal structures. The Peierls thermal conductivity ($\kappa_{\mathrm{p}}$) for all materials was evaluated as the average of 20 ensemble predictions. Magnetic materials, including those containing transition metals, were included in the screening. Although magnetic effects can affect lattice thermal conductivity in three-dimensional systems with Curie temperatures close to room temperature[110], they are generally secondary to phonon–phonon scattering because of the abundance and strength of phonon–phonon interactions[96,111]. Each model was trained on 3,000 anharmonic phonon data points, divided into 2,400 for training, 300 for validation, and 300 for testing. Following the screening, phonon properties, including $\kappa_\mathrm{p}$, were computed for 169 selected materials (148 with the highest $\kappa_\mathrm{p}$ and 21 with the lowest) using the auto-kappa workflow.

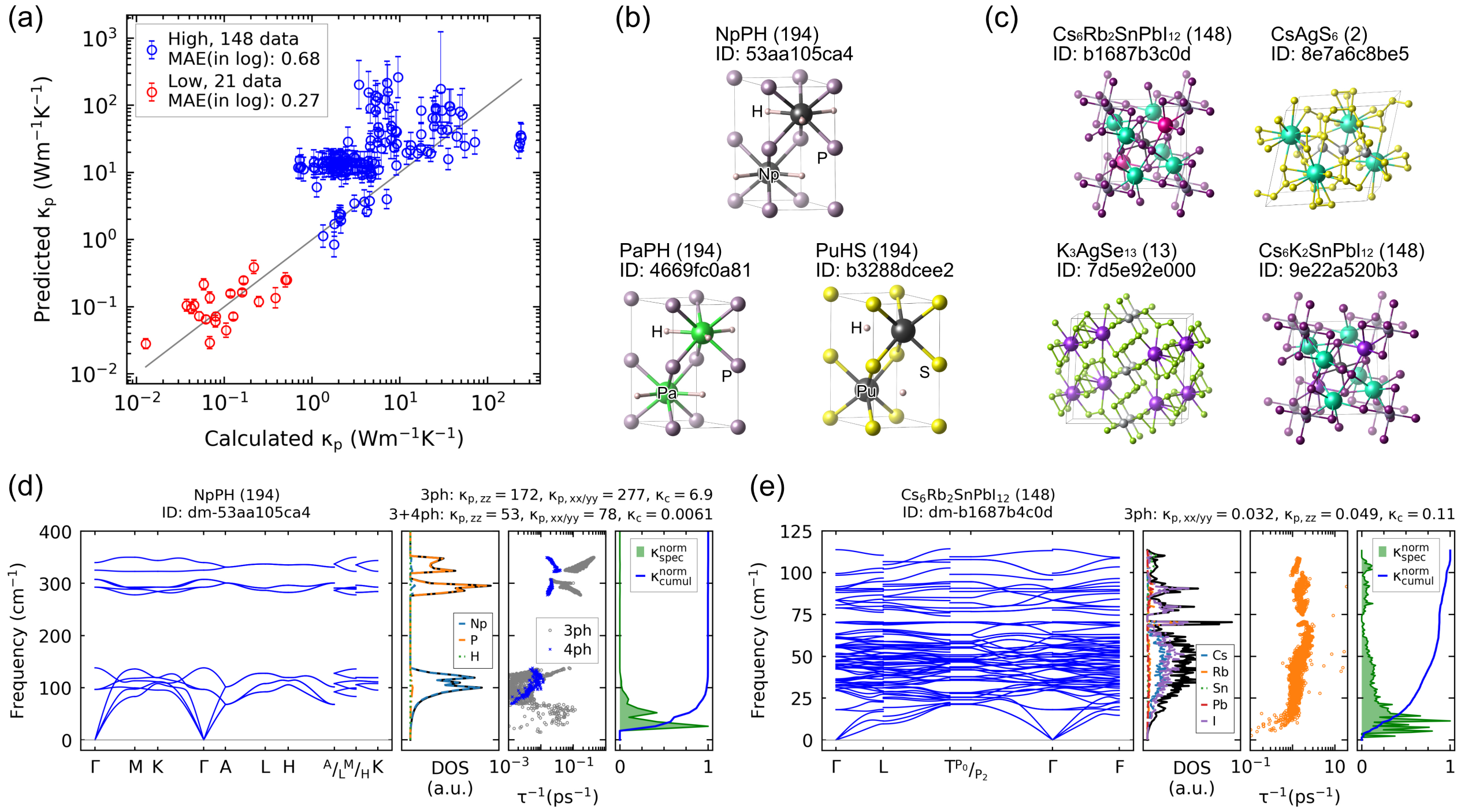

Fig. 4: Screening of high- and low-thermal conductivity materials from the GNoME database[14], which includes approximately[14] 381,000 novel structures. (a) Parity plot comparing predicted and calculated values of Peierls thermal conductivity ($\kappa_{\mathrm{p}}$). Blue and red markers represent materials predicted to exhibit high and low thermal conductivity, respectively, using models trained on the constructed the Phonix database. Error bars indicate the 90% confidence interval from 20 ensemble predictions. The solid line denotes the parity line. (b) and (c) display[136] crystal structures with $\kappa_{\mathrm{p}}^{\mathrm{3ph}} > 200\ \mathrm{Wm^{-1}K^{-1}}$ and the four lowest-$\kappa$ structures. For each material, the chemical formula, space group number (in parentheses), and GNoME database ID are provided (d) and (e) present phonon properties of hexagonal NpPH and trigonal $Cs_6Rb_2SnPbI_{12}$, which exhibit the highest ($\kappa_{\mathrm{lat}}^{\mathrm{3ph(3+4ph)}} \approx 280\ (80)\ \mathrm{Wm^{-1}K^{-1}}$) and lowest ($\kappa_{\mathrm{lat}} \approx 0.15\ \mathrm{Wm^{-1}K^{-1}}$) lattice thermal conductivities ($\kappa_{\mathrm{lat}} = \kappa_{\mathrm{p}} + \kappa_{\mathrm{c}}$), respectively. In the case of high-κ materials, both three-phonon (3ph) and four-phonon (4ph) scattering were taken into account. The panels include phonon dispersion, total and partial DOS, phonon lifetime ($\tau$), spectral (green) and cumulative (blue) Peierls thermal conductivity for each, as well as labels such as the chemical formula, space group (in parentheses), material ID, and lattice thermal conductivities ($\kappa_p$ and $\kappa_c$) along different directions in units of $\mathrm{Wm^{-1}K^{-1}}$. While the maximum phonon frequency of the high-$\kappa$ material in (d) exceeds 1,000 $\mathrm{cm^{-1}}$, properties are shown up to 400 $\mathrm{cm^{-1}}$. Full-range phonon properties are available in Supplementary Fig. S11. Spectral and cumulative thermal conductivity are normalized by the maximum and total Peierls conductivities, respectively.

An analysis of the validation results for the screened materials revealed several insights regarding prediction accuracy, as shown in Fig. 4(a). The predicted $\kappa_\mathrm{p}$ values for low-thermal-conductivity materials in the GNoME database showed accuracy comparable to that of the full dataset (MAE: 0.27 for $\log_{10}\kappa_\mathrm{p}$), with low variability in the predictions, as illustrated in Fig. 3(a). In contrast, the prediction accuracy for high-$\kappa_\mathrm{p}$ materials was notably lower (MAE: 0.68), and the predictions exhibited greater variability. Although definitive conclusions are limited by the relatively small number of computed data points, these results suggest that high-κ predictions are more challenging. From a machine learning standpoint, this difficulty likely stems from the simpler structural characteristics of high-$\kappa$ materials, which typically contain fewer atoms and atomic species in their primitive cells. Consequently, these materials offer less structural information for learning compared to low-$\kappa$ materials, which often have complex frameworks, such as skutterudites and clathrates[23,112,113]. Predicting material properties from such sparse structural information is inherently more difficult. From a physical perspective, accurately estimating high-$\kappa$ values demands rigorous treatment of anharmonic phonon interactions and highly converged computational parameters, such as dense $\mathbf{q}$-point meshes, since even small errors in force constants can significantly impact the results. Nonetheless, the predicted candidates remain promising for high-$\kappa$ applications.

By screening materials with high and low κ, we identified three compounds with $\kappa_\mathrm{lat}^\mathrm{3ph} > 200\ \mathrm{Wm^{-1}K^{-1}}$ and nine with $\kappa_\mathrm{lat}^\mathrm{3ph} < 0.2\ \mathrm{Wm^{-1}K^{-1}}$, as shown in Supplementary Fig. S7, where the superscript "3ph" denotes three phonon scattering. Among the predicted materials, the highest and lowest calculated lattice thermal conductivities ($\kappa_\mathrm{lat}^\mathrm{3ph} = \kappa_\mathrm{p}^\mathrm{3ph} + \kappa_\mathrm{c}^\mathrm{3ph}$) were 284 $\mathrm{Wm^{-1}K^{-1}}$ for the $xx$ and $yy$ components of hexagonal NpPH, and 0.14 $\mathrm{Wm^{-1}K^{-1}}$ for trigonal $Cs_6Rb_2SnPbI_{12}$, respectively, where $\left(\kappa_{\mathrm{p},\{xx/yy\}}, \kappa_\mathrm{c}\right) = (0.031, 0.11)$ $\mathrm{Wm^{-1}K^{-1}}$. Although we did not find materials that surpassed known record values, the results highlight the potential for future discovery of record-breaking compounds. Importantly, the identified candidates offer valuable insights into the structural and compositional characteristics of both high- and low-$\kappa$ materials. Discovering materials at the extremes of thermal conductivity is inherently challenging, as machine learning models typically excel at interpolation but struggle with extrapolation[114–116]. Therefore, further advancement in automated high-throughput calculations will be critical for identifying such extreme materials in future studies.

In the three-phonon calculations, high thermal conductivity values ($\gtrsim 200\ \mathrm{Wm^{-1}K^{-1}}$) were observed in hydrogen-containing hexagonal ternary compounds belonging to space group 194 ($P6_3/mmc$), such as NpPH ($\kappa_{p,zz}^\mathrm{3ph} = 172$, $\kappa_{p,xx/yy}^\mathrm{3ph} = 277$, $\kappa_c^\mathrm{3ph} = 6.9\ \mathrm{Wm^{-1}K^{-1}}$), PaPH ($\kappa_{p,zz}^\mathrm{3ph} = 173$, $\kappa_{p,xx/yy}^\mathrm{3ph} = 264$, $\kappa_c = 0.0037\ \mathrm{Wm^{-1}K^{-1}}$), and PuHS ($\kappa_{p,xx/yy/zz} = 216$, $\kappa_c = 0.012\ \mathrm{Wm^{-1}K^{-1}}$), as shown in Fig. 4(b) and Supplementary Fig. S11(a). When four-phonon scattering is taken into account, the thermal conductivity is reduced to $\kappa_{p,xx/yy}^\mathrm{3+4ph}$ = 78, 59, and 51 $\mathrm{Wm^{-1}K^{-1}}$ for NpPH, PaPH, and PuHS, respectively. The origin of their relatively high thermal conductivity nevertheless remains to be elucidated. These materials are characterized by heavy atoms surrounded by light atoms, including hydrogen. The phonon dispersion and DOS in Fig. 4(d) clearly show that phonon modes associated with heavy atoms (Np) and those associated with light atoms (P and H) are completely separated into different frequency ranges: modes of heavy atoms appearing at low frequencies ($< 200\ \mathrm{cm}^{-1}$) and those of light atoms appearing at high frequencies. This complete separation of phonon modes by different atomic species in energy space is expected to suppress anharmonic interactions between phonon modes within their respective frequency ranges, similar to other high-κ materials such as BAs[117,118]. Consequently, the phonon lifetimes of acoustic modes primarily composed of heavy atoms remain long, contributing dominantly to the overall heat transport, as shown in the last two panels of Fig. 4(d). In contrast, the crystal structures of low-$\kappa$ materials are significantly more complex, as illustrated by the examples in Fig. 4(c) such as $Cs_6Rb_2SnPBI_{12}$ ($\kappa_{p,zz} = 0.049$, $\kappa_{p,xx/yy} = 0.032$, $\kappa_c = 0.11\ \mathrm{Wm^{-1}K^{-1}}$), $CsAgS_6$ ($\kappa_{p,xx/yy/zz} = 0.013$, $\kappa_c = 0.141\ \mathrm{Wm^{-1}K^{-1}}$), $K_3AgSe_{13}$ ($\kappa_{p,xx/zz} = 0.030$, $\kappa_{p,yy} = 0.048$, $\kappa_c = 0.17\ \mathrm{Wm^{-1}K^{-1}}$), and $Cs_6K_2SnPbI_{12}$ ($\kappa_{p,zz} = 0.057$, $\kappa_{p,xx/yy} = 0.039$, $\kappa_c = 0.11\ \mathrm{Wm^{-1}K^{-1}}$). Notably, six of the nine discovered low-$\kappa$ materials contain cesium, whose alloy (α-$CsPbBr_3$) is known for its intrinsically low thermal conductivity [51]. In these low-$\kappa$ materials, phonon modes—formed by a mixture of atomic species—are distributed across a wide frequency range, as illustrated in Fig. 4(e) and Supplementary Fig. S11(b), in stark contrast to the more localized mode behavior seen in high-$\kappa$ materials. Although several attempts have been made to synthesize related materials, including actinide hydrides[119–121], the compounds identified in this screening—particularly those with high $\kappa_{\mathrm{lat}}$—may present significant challenges for experimental synthesis. Nevertheless, the above discussion provides concrete insight into the synthesis of highly thermally conductive materials. For example, realizing similar phenomena with transition metals, rather than actinides, could enable high thermal conductivity in compounds that are more amenable to experimental synthesis.

In conclusion, we developed an automated software package, auto-kappa, and constructed a large-scale first-principles database for anharmonic phonon interactions (Phonix), encompassing more than 6,000 materials with diverse crystal structures. Using this database, we demonstrated a clear scaling law linking dataset size to predictive performance for key anharmonic phonon properties, including lattice and spectral thermal conductivities. Furthermore, by screening a vast crystal structure database, we identified promising candidates for both high and low thermal conductivity applications. Although future improvements—such as the inclusion of higher-order anharmonic effects like four-phonon scattering and phonon renormalization—are necessary for more accurate assessments, this study establishes a strong foundation for data-driven discovery of thermofunctional materials with wide-ranging technological relevance, including applications in superconductivity, spintronics, and beyond.

## METHODS

### Automated Workflow for Anharmonic Phonon Calculations

Phonon calculations based on first-principles methods involve a considerably more complex workflow than typical calculations of total energy, electronic band structures, or electronic conductivity within the constant relaxation time approximation. To facilitate the construction of an anharmonic phonon property database, we developed auto-kappa, a Python-based automation software for first-principles analysis of anharmonic phonon properties. Auto-kappa streamlines the intricate workflow—illustrated in Fig. 1(a)—for computing anharmonic phonon properties by integrating the Vienna Ab Initio Simulation Package (VASP)[86] for electronic structure calculations and the phonon analysis software ALAMODE[45].

Through automated calculations, the auto-kappa software utilizes various existing libraries and packages in addition to VASP (≥ 6.3.2) and ALAMODE (versions 1.4–1.5). Crystal structures were handled using the Atomic Simulation Environment (ASE)[122] (≥ 3.22) and Pymatgen[7] (≥ 2023.8.10). Symmetry operations were performed using Spglib[123] (≥ 2.3.1), Pymatgen[7], and modules from Phonopy[52,124] (≥ 2.20). VASP calculations, including input file generation and job submission, were managed using ASE and the Custodian package[7] (≥ 2023.10.9). The phonon dispersion path was determined using the SeeK-path library[123,125].

The integration of various libraries—such as those listed above—enables researchers to perform first-principles phonon calculations with significantly reduced manual effort. Using auto-kappa, the database was generated through the following procedure, which follows the workflow illustrated in Fig. 1(a).

### i) Symmetry analysis of the given crystal structure

The primitive, conventional, and supercells of the input crystal structure were first determined. The conventional cell was selected to have a compact shape while maintaining resemblance to a regular hexahedron. The supercell was then generated from the conventional cell, with a target of maximizing the number of atoms (up to a limit of 150 atoms) while maintaining geometric similarity to a regular hexahedron. The resulting supercell was used for force calculations required for both harmonic and cubic force constants—steps iv and vi, respectively. However, when imaginary frequencies appeared, larger supercells were employed specifically for the harmonic force constant calculations.

### ii) Structure optimization

The accurate calculation of atomic forces using supercells in a later step is crucial for obtaining an reliable phonon analysis. Therefore, the shape and atomic positions in the crystal structure were carefully optimized through a rigorous procedure. Although both primitive and conventional cells can be used for this purpose, we chose the conventional cell to ensure consistency in the basis wavefunctions with those used in the supercell-based phonon calculations. While the primitive cell offers computational efficiency and better symmetry preservation, the conventional cell provides a more consistent basis set across all simulation steps.

The structure optimization was performed in three steps: two successive full relaxations—allowing for optimization of both the cell shape/volume and atomic positions—followed by a final atomic relaxation with the cell shape and volume fixed. Because changes in the cell can affect the optimal basis set of wavefunctions, performing two full relaxations helps mitigate the impact of basis fluctuations. Once the cell shape and size were determined, the atomic positions were further relaxed in a single-step calculation.

### iii) Calculation of Born effective charges

The Born effective charges were calculated using a first-principles approach to apply non-analytical corrections in subsequent phonon analyses. For harmonic phonon properties, such as phonon dispersion and DOS, the non-analytic correction was initially applied using the mixed-space approach[126]. This correction primarily affects the splitting between longitudinal optical (LO) and transverse optical (TO) modes (LO–TO splitting), but in some cases, it also influences the phonon stability of certain materials. When imaginary phonon frequencies were observed, the method for applying the non-analytic correction was modified—first by using the damping method[127] and, if necessary, switching to the Ewald method[128].

### iv) Calculation of harmonic force constants

Harmonic interatomic force constants were calculated using the finite-displacement method (also known as the brute-force method), in which atomic displacement patterns were generated in a supercell, and the resulting atomic forces were computed for each pattern. For these calculations, a single atom was displaced within the supercell, and the displacement patterns were determined based on crystal symmetry. The number of displacement patterns required for harmonic force constants is relatively small compared to those needed for higher-order force constants, allowing the finite-displacement method to be directly applied. The displacement magnitude was set to a small value (0.01 Å) to minimize the influence of anharmonic effects. Harmonic force constants were then obtained using a least-squares fitting procedure. If the fitting error exceeded 10%, the data were excluded from the analyses presented in this paper. No cutoff distance was imposed on the harmonic force constants in order to account for all possible atomic interactions within the supercell.

To ensure accurate force calculations within the first-principles framework, it is important to evaluate the nonlocal part of the pseudopotential in reciprocal space rather than in real space. While using projector operators in real space can reduce computational cost for large supercells, it introduces aliasing errors due to wavefunction projection. Therefore, in our developed software, projector operators are consistently evaluated in reciprocal space by setting '$\mathrm{LREAL} = \mathrm{FALSE}$' in the VASP calculations.

### v) Analysis for harmonic phonon properties

Using harmonic force constants, harmonic phonon properties—including phonon dispersion and DOS—were calculated. As described in the section on the Born effective charge, different approaches were applied to include non-analytic corrections when necessary to eliminate imaginary frequencies. For the DOS calculation, the reciprocal space mesh density for the phonon wavevector ($\mathbf{q}$-mesh) was set to 1500 q-points per reciprocal atom ($\mathrm{q\text{-}points{\cdot}Å^3/atom}$). For example, the $\mathbf{q}$-mesh for diamond-structured silicon was set to $21 \times 21 \times 21$.

### vi) Calculation of cubic force constants

If the structure exhibited no imaginary frequencies, the calculation of cubic force constants was performed following the harmonic phonon property analysis. To obtain cubic force constants, the finite-displacement method typically requires a significantly larger number of displacement patterns—on average, approximately 100 times more than those needed for harmonic force constants. Therefore, a cutoff distance was imposed for the cubic force constants, which was set to the larger of 4.3 Å and the third-shortest interatomic distance. Additionally, while the finite-displacement and least-squares methods were used when the number of required displacement patterns was below a predefined threshold (set to 100 patterns), the least absolute shrinkage and selection operator (LASSO) regression[129] was employed to estimate cubic force constants from randomly generated displacement patterns. The harmonic force constants were fixed to the values obtained from the previous calculation (step iv) during the LASSO regression. If the fitting error for the least-squares method or the residual force for the LASSO regression exceeded 10%, the data was excluded from the discussion, as was done for harmonic force constants.

The number of generated random displacement patterns was determined using the formula $N_{\mathrm{pattern}}^{\mathrm{rand}} = \alpha N_{\mathrm{FC3}}/N_{\mathrm{atom}}^{\mathrm{sc}}$, where $N_{\mathrm{FC3}}$ is the number of unique cubic force constants, $N_{\mathrm{atom}}^{\mathrm{sc}}$ is the number of atoms in the supercell, and $\alpha$ is a coefficient greater than $1/3$; in this study, it was set to $1.0$. To generate a random displacement pattern, a random displacement was applied to each atom. The displacement magnitude for cubic calculations was set to 0.01 or 0.03 Å per atom for both the finite-displacement method and the LASSO approach, which is larger than the value used for harmonic calculations.

### vii) Analysis for anharmonic phonon properties

Using the cubic force constants obtained in the previous step, we analyzed anharmonic phonon properties. To assess convergence with respect to the $\mathbf{q}$-mesh size, the $\mathbf{q}$-mesh density was varied from 500 to 1000 to 1500 $\mathrm{q\text{-}points{\cdot}Å^3/atom}$. The effect of three-phonon scattering was estimated by solving the phonon transport Boltzmann equation under the relaxation time approximation. Phonon scattering by natural isotopes was also considered and incorporated using Matthiessen's rule. Finally, various anharmonic phonon properties were obtained, including mode-dependent lifetimes; spectral and cumulative thermal conductivities ($\kappa_{\mathrm{spec}}$ and $\kappa_{\mathrm{cumul}}$) as functions of frequency and mean free path; and temperature-dependent thermal conductivities for both Peierls ($\kappa_{\mathrm{p}}$) and coherence ($\kappa_{\mathrm{c}}$)[59] contributions, as illustrated in Fig. 1(b). For details, please refer to Section I of the Supplementary Information.

### viii) Strict structure optimization

If imaginary frequencies were observed in the harmonic phonon analysis during process (iv), a strict structural optimization was performed. In this step, the volume of the crystal structure was modified by applying hydrostatic strain, and the corresponding structural energies were calculated. After evaluating energies at different volumes, the Birch-Murnaghan equation of state[130,131] was used to determine the volume that minimized the structural energy. Once the newly optimized structure was obtained, the procedure was restarted from process (iii).

### ix) Use of larger supercell for harmonic force constants

If the strictly optimized structure still exhibited imaginary frequencies, a larger supercell was used for calculating harmonic force constants. The maximum limit for this second harmonic force constant analysis was set to 200 atoms—an increase of 50 atoms from the original setting. If this step successfully eliminated imaginary frequencies, cubic force constants were then calculated. While a larger supercell was used for harmonic force constants in this case, the original supercell size (fewer than 150 atoms) was retained for estimating cubic force constants. The harmonic force constants obtained using the original supercell were kept fixed during the estimation of cubic force constants.

### x) Phonon renormalization

The process for phonon renormalization using self-consistent phonon (SCP) theory[55,56] was also implemented in auto-kappa, although this process was not performed in the present study. Using the SCP approach, temperature-dependent effective harmonic force constants can be calculated by incorporating the effects of phonon renormalization due to the fourth-order potential. Phonon renormalization can eliminate imaginary frequencies in certain cases [56,113] , and should also be considered for accurately estimating low thermal conductivity.

For all first-principles simulations described above, the following conditions were applied. The $\mathbf{k}$-mesh was determined by $N_i = \max\left[1, \mathrm{int}(l_k \cdot |\mathbf{b}_i|)\right]$, following the method recommended by VASP. Here, $l_k$ is a length scale that determines the number of subdivisions along each reciprocal lattice direction and is set to 20 Å, and $\mathbf{b}_i$ is the reciprocal lattice vector along the $i$ direction ($i = k_x, k_y, k_z$). The Γ-centered scheme was used to generate the $\mathbf{k}$-mesh. The Perdew-Burke-Ernzerhof exchange-correlation functional revised for solids (PBEsol)[132] with the projector augmented wave (PAW) potential[133,134] was employed. The cutoff energy for VASP calculations was set to 1.3 times the recommended value provided in the VASP pseudopotential files.

## Machine Learning Prediction of Phonon Properties

We employed the crystal graph convolutional neural network (CGCNN)[103] to predict the Peierls conductivity ($\kappa_{\mathrm{p}}$) and the graph neural network based on the Euclidean neural network (e3nn)[104,106] to predict spectral functions and cumulative Peierls conductivity ($\kappa_{\mathrm{cumul}}$) as a function of the phonon mean free path (Λ). In both graph neural network approaches, nodes and edges correspond to atoms and bonds within the crystal, respectively.

The node descriptors in CGCNN consist of a one-hot encodings of nine atomic properties, including group number, period number, electronegativity, and covalent radius, as also described in the main text. In contrast, the e3nn approach employs a simpler node descriptor: a 118-dimensional mass-weighted one-hot encoding based solely on atomic species and their masses. For edge descriptors, CGCNN utilizes a 10-dimensional encoding based on interatomic distances categorized into discrete intervals, whereas e3nn encodes edges using full three-dimensional relative position vectors between neighboring atoms, explicitly capturing both geometric and directional information. The cutoff bond lengths were set to 6.0 Å and 4.3 Å for CGCNN and e3nn, respectively.

Both graph neural networks employ multiple convolutional layers to update atomic features by aggregating local atomic environments. In CGCNN, three graph convolutional layers sequentially update node features using information from up to 12 nearest neighbors. A pooling layer aggregates atomic-level features into a global crystal representation, which is subsequently mapped to scalar material properties through fully connected layers. The e3nn approach utilizes convolutional layers constructed from spherical harmonics and learnable radial basis functions, designed to ensure equivariance under rotations, translations, and inversions. The network typically includes two equivariant convolutional layers followed by gated nonlinearity blocks tailored for tensorial data. After convolution and activation, atomic features are aggregated to form a global descriptor, which is directly mapped to continuous spectral functions, namely the cumulative ($\kappa_{\mathrm{cumul}}^{\mathrm{norm}}$) and spectral ($\kappa_{\mathrm{spec}}^{\mathrm{norm}}$) thermal conductivities.

The neural networks were trained using the Adam optimizer[135]. For CGCNN, the learning rate was set to 0.0001, and early stopping was applied with a patience of 50 epochs. While the prediction performance of CGCNN was relatively insensitive to hyperparameter choices, the hyperparameters for the e3nn approach—particularly the learning rate—were carefully tuned. The initial learning rate was set to $5.0/N_{\mathrm{all}}$ and decayed by a factor of 0.95 per epoch until it reached a minimum of $1.5/N_{\mathrm{all}}$, where $N_{\mathrm{all}}$ denotes the total number of data points, including training, validation, and test sets. Early stopping was applied with a patience of 100 epochs during e3nn training.

For training in both cases, the simulation dataset was divided into three parts: training data (80%), validation data (10%), and test data (10%). The training data were used to develop the prediction model, while the validation data were used to tune hyperparameters and prevent overfitting. The test data were employed to evaluate the prediction error. The size of the simulation dataset was varied from 100 to the full dataset (approximately 5,000 samples), and 20 ensembles were generated to assess the fluctuation in prediction performance. Log scaling and normalization were applied to the target values for $\kappa_{\mathrm{p}}$ and $\kappa_{\mathrm{cumul}}(\Lambda)$, respectively. Therefore, if the absolute value of $\kappa_{\mathrm{cumul}}(\Lambda)$ is required, it can be reconstructed by combining the two predictions.

For the prediction of $\kappa_{\mathrm{cumul}}$, the data were prepared over a range from 1 nm to 100 μm, sampled at 51 logarithmically spaced points. The performance of the prediction model was evaluated using the mean absolute error (MAE). The MAE for each material was computed as $|\kappa_{\mathrm{p}}^{\mathrm{calc}} - \kappa_{\mathrm{p}}^{\mathrm{pred}}|$ for $\kappa_{\mathrm{p}}$, and as $\sum_{\Lambda} |\kappa_{\mathrm{cumul}}^{\mathrm{calc}}(\Lambda) - \kappa_{\mathrm{cumul}}^{\mathrm{pred}}(\Lambda)|$ for $\kappa_{\mathrm{cumul}}(\Lambda)$, where the superscripts "calc" and "pred" refer to the calculated and predicted values, respectively. The final MAE was obtained by averaging over the entire test dataset. After calculating the MAE for various training data sizes ($N_{\mathrm{train}}$), the scaling law was determined by fitting the relationship using the function (MAE) $= (N_{\mathrm{c}}/N_{\mathrm{train}})^{\alpha}$ $(N_c, \alpha > 0)$, where $N_{\mathrm{c}}$ is a constant and $\alpha$ is the scaling factor indicating how efficiently increasing the data size improves prediction accuracy.

## DATA AVAILABILITY

The dataset used for machine learning prediction, along with the Python scripts employed in this study, is available in the GitHub repository at https://github.com/masato1122/phonon_e3nn. Phonix—a database for anharmonic phonon interactions—will be made available on ARIM-mdx at https://phonix.org.

## CODE AVAILABILITY

Software for the automated calculation of anharmonic phonon properties (auto-kappa), as well as for the machine learning prediction of these properties, will be made available in the GitHub repository at

https://github.com/masato1122/auto-kappa.

## ACKNOWLEDGEMENTS

The authors thank Chris Dames and Ying Sun for co-organizing the Workshop "Thermal Transport, Materials Informatics, and Quantum Computing" supported by National Science Foundation (NSF) and Japan Science and Technology Agency (JST), where this project was conceptualized. The authors also thank C. Wolverton, A. Togo, K. Esfarjani, and M. Kawamura for fruitful discussions. Numerical calculations were performed using the following supercomputers through the HPCI System Research Project (Project IDs: hp220151, jh230065, and hp240194): Grand Chariot at the Information Initiative Center, Hokkaido University; OCTOPUS and SQUID at the D3 Center, Osaka University; Oakbridge-CX and Wisteria/BDEC-01 at the Supercomputing Division, Information Technology Center, The University of Tokyo; and AOBA-B at the Cyberscience Center, Tohoku University. Additional resources were provided by the Supercomputer Center, Institute for Solid State Physics, The University of Tokyo, and MASAMUNE-IMR at the Center for Computational Materials Science, Institute for Materials Research, Tohoku University. This work was partially supported by JSPS KAKENHI Grants No. 24K07354 and No. 22H04950 from the Japan Society for the Promotion of Science (JSPS), CREST Grants No. JPMJCR19I2 and No. JPMJCR21O2 from the Japan Science and Technology Agency (JST), and a grant-in-aid from the Thermal and Electric Energy Technology Foundation. K.H. acknowledges funding from the MAT-GDT Program at A*STAR via the AME Programmatic Fund by the Agency for Science, Technology and Research under Grant No. M24N4b0034. L. L. acknowledges supported for vibrational property calculations and database discussions from the U.S. Department of Energy, Office of Science, Office of Basic Energy Sciences, Material Sciences and Engineering Division. T.D. acknowledges the financial support from National Natural Science Foundation of China (Grant No. 62204218) and Leading Innovative and Entrepreneur Team Introduction Program of Hangzhou (No. TD2022012), and computational resources from the National Supercomputer Center in Tianjin. P. T. acknowledges the financial support from the Catalan Government through the funding grant ACCIÓ-Eurecat (Project TRAÇA SMART-MAT).

## AUTHOR CONTRIBUTIONS

The project was conceptualized by T.L. and J.S. (together with Chris Dames and Ying Sun), and managed by M.O. and J.S. M.O., T.T., T.D., P.T., and Z.X. contributed to code development. M.O., T.D., P.T., Z.X., H.Z., and W.N. generated phonon property data through automated calculations. M.O., R.Y., and J.S. contributed to data analysis. M.O., M.H., T.S., R.Y., and J.S. contributed to the machine learning and database construction. M.O. and J.S. wrote the original manuscript, and all authors contributed to revising the manuscript.

## COMPETING INTERESTS

The authors declare no competing interests.

## ADDITIONAL INFORMATION

**Supplementary information**

The online version contains supplementary material available at ***.

**Correspondence** and requests for materials should be addressed to Masato Ohnishi (masato.ohnishi.ac@gmail.com) and Junichiro Shiomi (shiomi@photon.t.u-tokyo.ac.jp).

**Supplementary Information for**

**"Database and deep-learning scalability of anharmonic phonon properties by automated brute-force first-principles calculations"**

Masato Ohnishi[1,2,*] Tianqi Deng[3,4], Pol Torres[5], Zhihao Xu[6], Terumasa Tadano[7], Haoming Zhang[3,4], Wei Nong[8], Masatoshi Hanai[9], Zeyu Wang[10], Zhiting Tian[11], Ming Hu[12], Xiulin Ruan[13], Ryo Yoshida[2,14,15], Toyotaro Suzumura[9], Lucas Lindsay[16], Alan J. H. McGaughey[17], Tengfei Luo[6,18], Kedar Hippalgaonkar[8,19,20], and Junichiro Shiomi[1,2,10,21,*]

[1] Institute of Engineering Innovation, The University of Tokyo, Tokyo 113-0032, Japan

[2] The Institute of Statistical Mathematics, Research Organization of Information and Systems, Tachikawa, Tokyo 190-8562, Japan

[3] State Key Laboratory of Silicon and Advanced Semiconductor Materials, School of Materials Science and Engineering, Zhejiang University, Hangzhou 310027, China

[4] Key Laboratory of Power Semiconductor Materials and Devices of Zhejiang Province, Institute of Advanced Semiconductors, ZJU-Hangzhou Global Scientific and Technological Innovation Center, Zhejiang University, Hangzhou 311200, China

[5] Eurecat, Technology Centre of Catalonia, Unit of Applied Artificial Intelligence, Cerdanyola del Vallès, 08290, Spain

[6] Department of Aerospace and Mechanical Engineering, University of Notre Dame, Notre Dame, IN 46556, USA

[7] Research Center for Magnetic and Spintronic Materials, National Institute for Materials Science, Tsukuba 305-0047, Japan

[8] School of Materials Science and Engineering, Nanyang Technological University, Singapore 639798, Singapore

[9] Information Technology Center, The University of Tokyo, Tokyo 113-0032, Japan

[10] Department of Mechanical Engineering, The University of Tokyo, Tokyo 113-0032, Japan

[11] Sibley School of Mechanical and Aerospace Engineering, Cornell University, Ithaca, New York 14853, USA

[12] Department of Mechanical Engineering, University of South Carolina, Columbia, SC 29201, USA

[13] School of Mechanical Engineering and Birck Nanotechnology Center, Purdue University, West Lafayette, IN 47907, USA

[14] The Graduate University for Advanced Studies, SOKENDAI, Tachikawa, Tokyo, 190-8562, Japan

[15] Advanced General Intelligence for Science Program (AGIS), RIKEN-TRIP, Wako, Saitama 351-0198, Japan

[16] Materials Science and Technology Division, Oak Ridge National Laboratory, Oak Ridge, TN 37831, USA

[17] Department of Mechanical Engineering, Carnegie Mellon University, Pittsburgh, Pennsylvania 15213, USA

[18] Department of Chemical and Biomolecular Engineering, University of Notre Dame, Notre Dame, IN 46556, USA

[19] Institute of Materials Research and Engineering, Agency for Science Technology and Research, Innovis, Singapore 138634, Singapore

[20] Institute for Functional Intelligent Materials, National University of Singapore, Singapore 117544, Singapore

[21] RIKEN Center for Advanced Intelligence Project, Tokyo 103-0027, Japan

# I. FORMULATIONS FOR LATTICE THERMAL CONDUCTIVITY

Lattice thermal conductivity was calculated by solving Peierls-Boltzmann transport equation[1] for the particle-like contribution and the Wigner transport formula[2] for the wave-like contribution, both implemented in the ALAMODE package[3,4]. Considering the second-order perturbation within the relaxation approximation, the phonon linewidth due to the three-phonon scattering for phonon mode $q$ is derived as

$$\Gamma_q = \frac{\pi}{16N} \sum_{q_1,q_2} |V_3(-q,q_1,q_2)|^2 \left[(n_1+n_2+1)\delta(\omega_q-\omega_1-\omega_2) - 2(n_1-n_2)\delta(\omega_q-\omega_1+\omega_2)\right], \quad \text{(S1)}$$

where the subscripts ($i$ = 1, 2) denote phonon modes contributing to the scattering of the target mode $q$, $n_i$ is the Bose-Einstein distribution function, $\omega_i$ is the phonon frequency, $N$ is the number of $\boldsymbol{q}$ points, and $\pm q = (\pm\boldsymbol{q}, s)$ with $\boldsymbol{q}$ and $s$ being the wavevector and branch index, respectively. The three-phonon coupling matrix element $V_3$ is given by

$$V_3(q,q_1,q_2) = \left(\frac{\hbar}{\omega\omega_1\omega_2}\right)^{\frac{1}{2}} \times \sum_{\boldsymbol{R}_i l_i p_i} \psi^{p_0 p_1 p_2}_{0 l_0, \boldsymbol{R}_1 l_1, \boldsymbol{R}_2 l_2} \times \frac{\boldsymbol{e}^{p_0}_{l_0}(q)\boldsymbol{e}^{p_1}_{l_1}(q_1)\boldsymbol{e}^{p_2}_{l_2}(q_2)}{\sqrt{M_{l_0}M_{l_1}M_{l_2}}} \times \exp[i(\boldsymbol{q}\cdot\boldsymbol{R}_0 + \boldsymbol{q_1}\cdot\boldsymbol{R}_1 + \boldsymbol{q_2}\cdot\boldsymbol{R}_2)], \quad \text{(S2)}$$

where $\hbar$ is the reduced Planck constant, $\boldsymbol{R}_i$ is the position of the primitive cell, $l_i$ is the atom index, $p_i$ is the direction of the displacement of atom $l_i$, $M$ is the atomic mass, $\Psi$ is the cubic force constants, and $\boldsymbol{e}(q)$ is the eigenvector of the mode $q$. The phonon lifetime due to three phonon scattering $\tau_{\text{pp}}$ is given by $\tau_{\text{pp}}(q) = 1/(2\Gamma_q)$. The total phonon lifetime of mode $q$ is obtained with Matthiessen's rule: $\tau_q^{-1} = \tau_{q,\text{pp}}^{-1} + \tau_{q,\text{iso}}^{-1}$, where $\tau_{q,\text{iso}}^{-1}$ is the scattering rate due to isotopes[5]. Finally, the Peierls term is calculated as $\kappa_{\text{p}}^{\alpha\beta,\text{scp}}(T) = (NV)^{-1}\sum_q c_q(T) v_q^{\alpha}(T) v_q^{\beta}(T)\tau_q(T)$. Here, $V$ is the volume of the primitive unit cell, $c$ is the mode specific heat, $v$ is the group velocity, and $\alpha$ and $\beta$ are the Cartesian directions.

In the automated workflow, the coherence contribution ($\kappa_{\text{c}}$), corresponding to the nondiagonal terms of the heat flux operator, was also considered. Its contribution can be obtained as

$$\kappa_{\text{c}}^{\alpha\beta} = \frac{\hbar^2}{k_{\text{B}}T^2VN} \sum_{\boldsymbol{q}} \sum_{s_1 \neq s_2} \frac{\omega_1+\omega_2}{4} \text{V}_{12}^{\alpha}(\boldsymbol{q}) \text{V}_{21}^{\beta}(\boldsymbol{q}) \frac{\omega_1 n_1(n_1+1) + \omega_2 n_2(n_2+1)}{(\omega_1-\omega_2)^2 + (\Gamma_1+\Gamma_2)^2} (\Gamma_1+\Gamma_2), \quad \text{(S3)}$$

where $k_{\text{B}}$ is the Boltzmann constant. The subscripts (1 and 2) denote the phonon modes $(\boldsymbol{q}, s_1)$ and $(\boldsymbol{q}, s_2)$, respectively. The generalized group velocity operator V($q$) is given as

$$\text{V}_{12}(\boldsymbol{q}) = \frac{1}{\omega_1+\omega_2} \left\langle \boldsymbol{e}(q_1) \middle| \frac{\partial D(\boldsymbol{q})}{\partial \boldsymbol{q}} \middle| \boldsymbol{e}(q_2) \right\rangle, \quad \text{(S4)}$$

where $D(\boldsymbol{q})$ is the dynamical matrix.

# II. ELEMENTAL OCCURRENCE FREQUENCY

The elemental occurrence frequency in the materials for which anharmonic properties were calculated in *Phonix*, a database for anharmonic phonon interactions, is illustrated in the periodic table shown in Fig. S1. The number shown for each element represents the number of materials that include that element. For example, the most frequently occurring element, oxygen, is found in 2,732 out of the 7,038 materials in the current version of Phonix.

This figure shows that chalcogens (group 16) and alkali metals (group 1) are the two most frequently occurring groups of elements. It also indicates that elements from groups 1–2 and 11–17 are commonly present in the Phonix materials, whereas transition metals occur less frequently—likely due to the absence of magnetic materials in the current version of Phoinix.

Overall, Phonix contains a broad range of elements, which is important for improving data quality and enabling its application to diverse fields—not only for thermal properties, but also for various other material properties, including mechanical, electronic, electrical, optical, and magnetic characteristics.

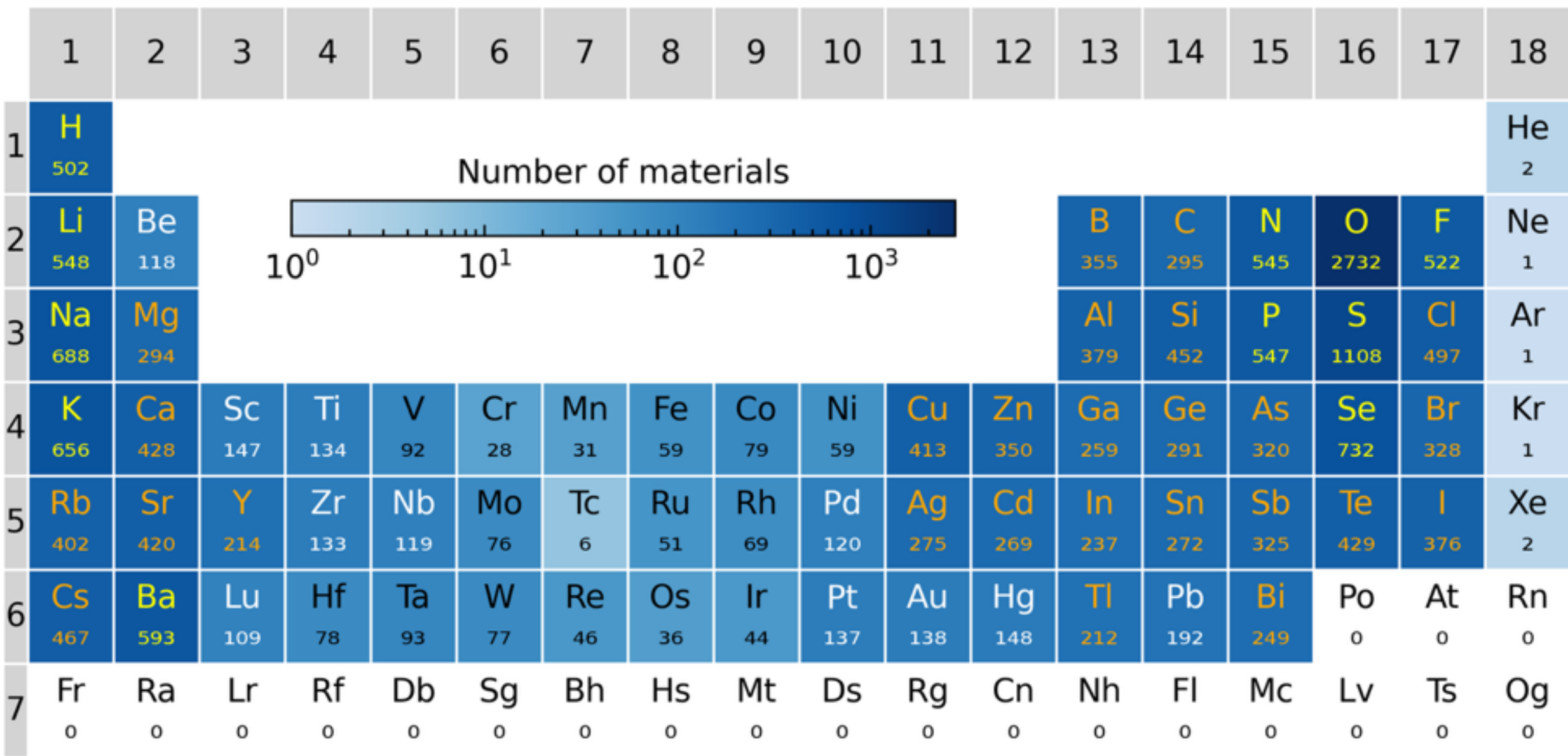

Fig. S1. Elemental occurrence frequency in the database for anharmonic phonon interactions (Phonix). The number in each cell indicates how frequently each element appears in the materials included in the Phonix database, with the magnitude visualized using a logarithmic blue color scale. Element symbols are colored according to their occurrence: yellow for elements appearing in more than 500 materials, orange for more than 200, white for more than 100, and black for all others.

## III. LIST OF MATERIALS WITH HIGH AND LOW THERMAL CONDUCTIVITIES

The target materials consist of non-metallic, non-magnetic materials from the Materials Project[6–8], as well as those from Phonondb[9], totaling 21,452 materials. To construct Phonix, we developed an automated calculation software package, auto-kappa, for anharmonic phonon properties. So far, we have calculated the anharmonic phonon properties of approximately 6,000 materials, while harmonic phonon properties have been computed for an even larger set of materials.

Among the calculated high-$\kappa$ materials, the following materials, including their polymorphs, are novel or have been rarely discussed as high-$\kappa$ materials: triclinic $Hg(BiS_2)_2$ ($\kappa_{p,\{xx,yy,zz\}}$ = 292, 2.5, and 943 $\mathrm{Wm^{-1}K^{-1}}$), cubic HC (306 $\mathrm{Wm^{-1}K^{-1}}$), cubic BiB (235 $\mathrm{Wm^{-1}K^{-1}}$), and trigonal $CsHoS_2$ ($\kappa_{p,\{zz,xx(yy)\}}$ = 22 and 657 $\mathrm{Wm^{-1}K^{-1}}$). Only 110 materials (1.6%) has a $\kappa_{\mathrm{lat}}$ exceeding 100 $\mathrm{Wm^{-1}K^{-1}}$, as shown in Table S1. On the other hand, discovering novel materials with low thermal conductivity appears to be much easier. In the Phonix database, nearly 2,000 materials (28%) have $\kappa_{\mathrm{lat}}$ below 1.0 $\mathrm{Wm^{-1}K^{-1}}$. Even when limited to binary compounds, 207 materials (3.0%) have a $\kappa_{\mathrm{lat}} \leq 1.0\ \mathrm{Wm^{-1}K^{-1}}$. Considering that the thermal conductivities of typical thermoelectric materials $Bi_2T_3$ and PbTe are 1.2[10] and 2.2[11–13] $\mathrm{Wm^{-1}K^{-1}}$, the constructed database provides an exciting opportunity to search for materials with low-$\kappa$ for applications such as thermal insulators and thermoelectric materials.

Phonon properties of the fifty-five high-$\kappa$ materials ($\kappa_{\mathrm{lat}} \geq 200\ \mathrm{Wm^{-1}K^{-1}}$) and fourteen low-$\kappa$ materials ($\kappa_{\mathrm{lat}} \leq 0.1\ \mathrm{Wm^{-1}K^{-1}}$) are shown in Fig. S2 and Fig. S3, respectively.

Table S1. Number of materials exhibiting high and low thermal conductivity in Phonix. The lattice thermal conductivity is evaluated as $\kappa_{\mathrm{lat}} = (\kappa_{xx} + \kappa_{yy} + \kappa_{zz})/3$ for all materials, including anisotropic ones, for simplicity.

| **High thermal conductivity** | | **Low thermal conductivity** | |
|---|---|---|---|
| **Range of $\kappa_{\mathrm{lat}}$ ($\mathrm{Wm^{-1}K^{-1}}$)** | **Number of materials** | **Range of $\kappa_{\mathrm{lat}}$ ($\mathrm{Wm^{-1}K^{-1}}$)** | **Number of materials** |
| >1,000 | 10 (0.14%) | <0.1 | 14 (0.20%) |
| >500 | 24 (0.35%) | <0.2 | 206 (3.0%) |
| >200 | 65 (0.92%) | <0.5 | 1036 (15%) |
| >100 | 110 (1.6%) | <1.0 | 1,912 (28%)<br>(207 for binary) |

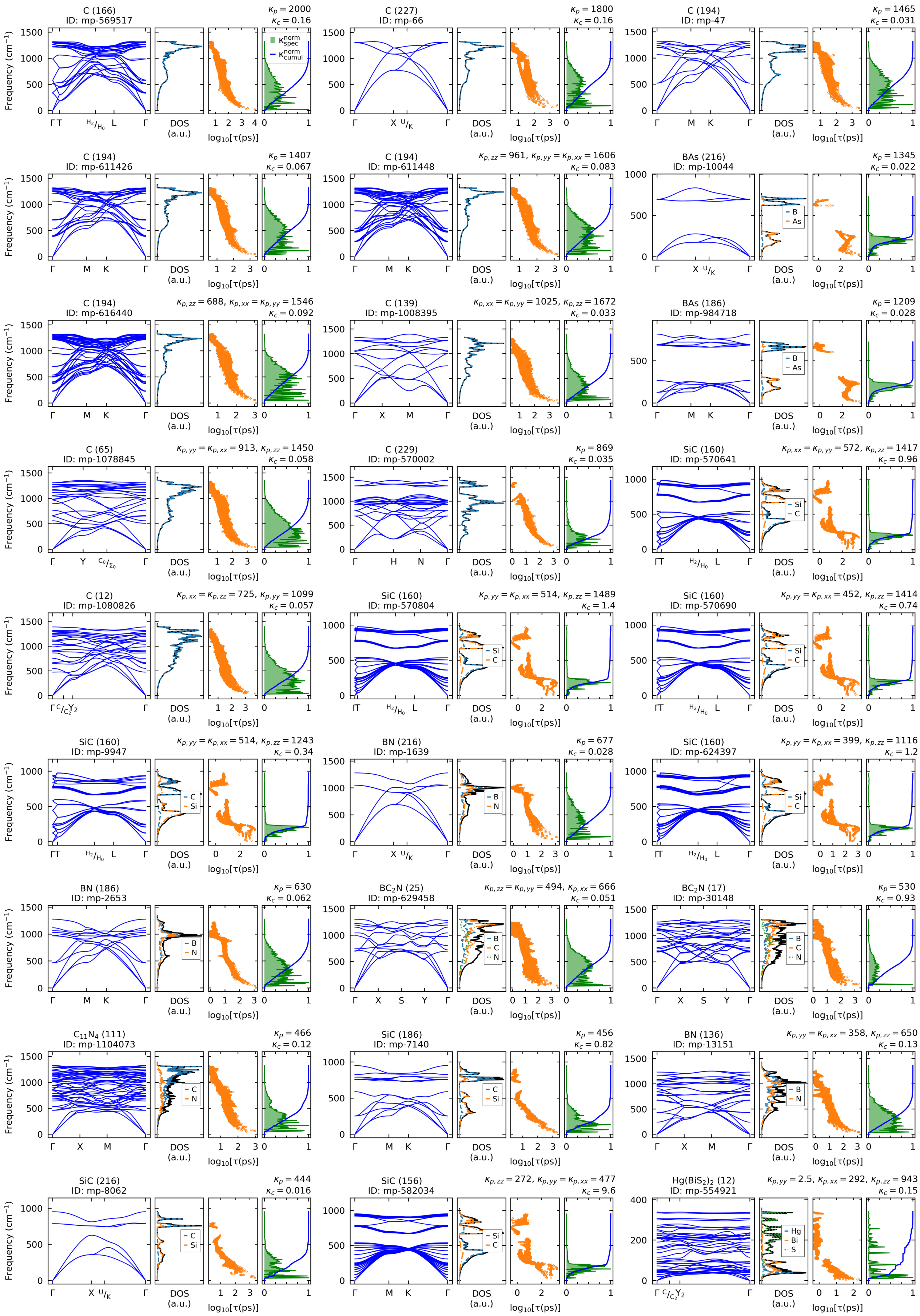

Fig. S2 (1/2). List of phonon properties for materials with high thermal conductivity

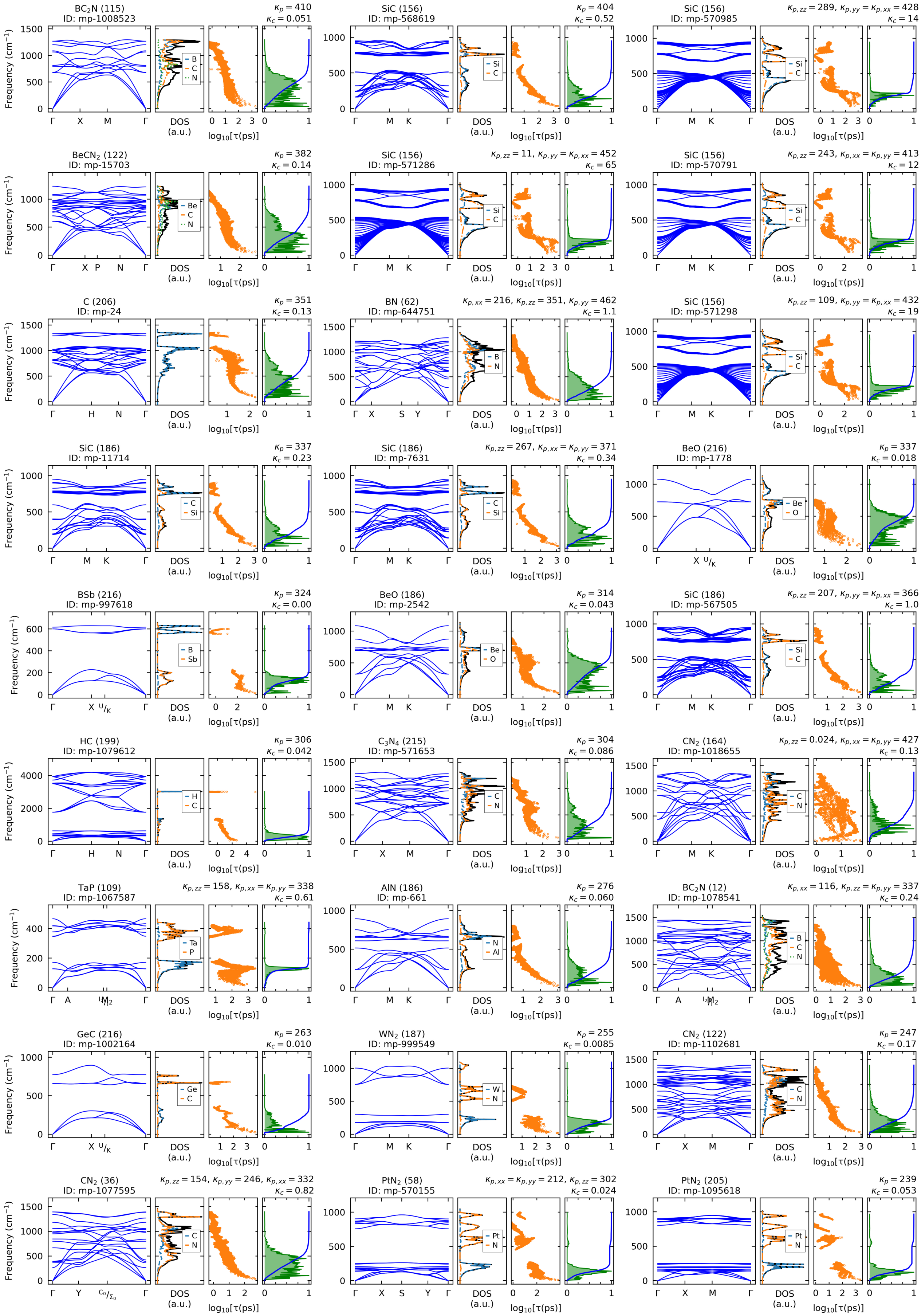

Fig. S2 (2/2). List of phonon properties for materials with high thermal conductivity.

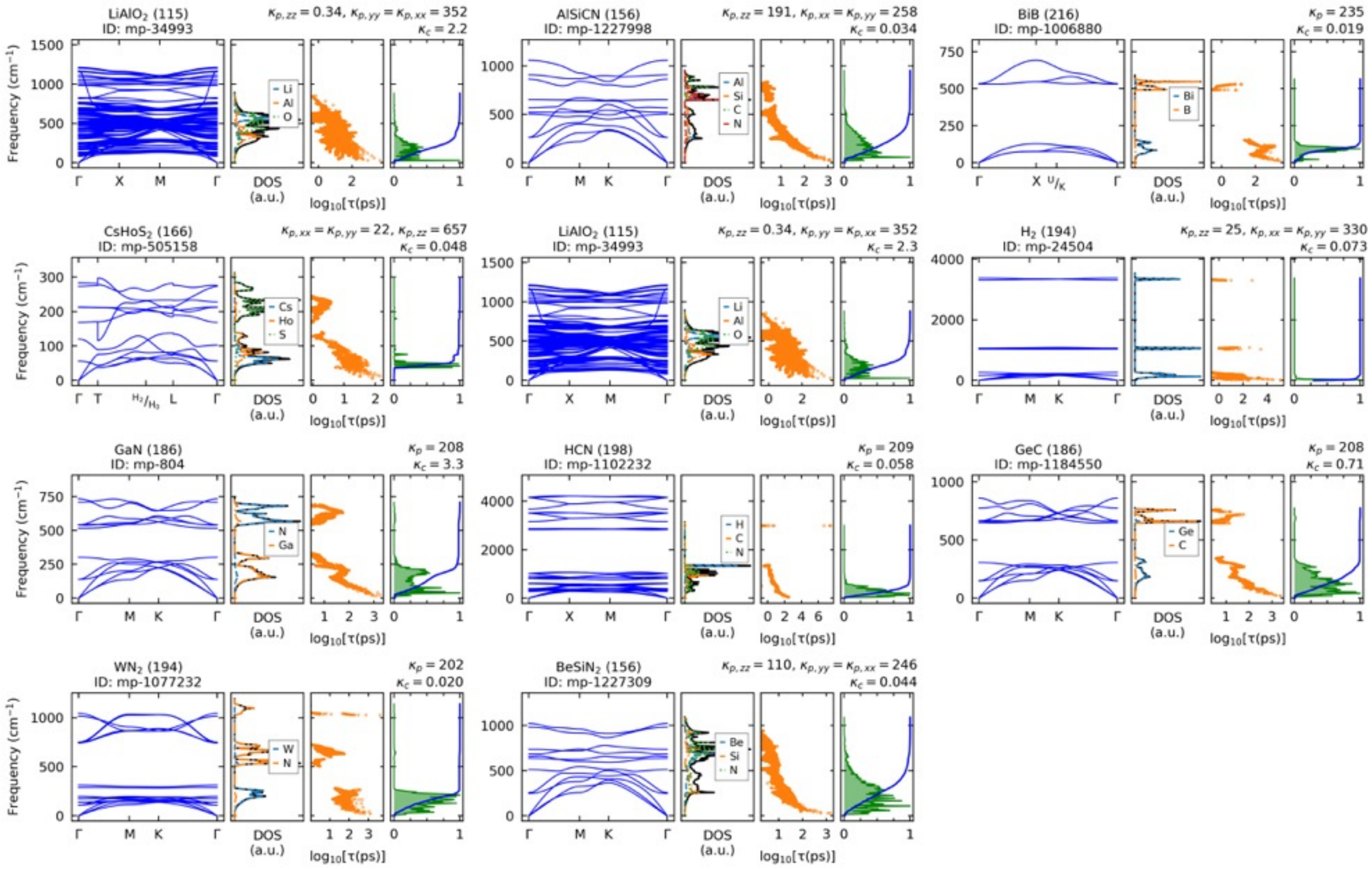

Fig. S2. List of phonon properties for materials with high thermal conductivity ($> 200$ Wm$^{-1}$K$^{-1}$), ordered by magnitude. Each panel includes phonon dispersion, density of states (DOS), phonon lifetime (in logarithmic scale), and spectral (green) and cumulative (blue) thermal conductivity for Peierls component ($\kappa_p$), which are normalized by their respective maximum and total values. The chemical formula, space group number (in parentheses), and Materials Project ID are shown in the top-left corner, while the magnitudes of the Peierls ($\kappa_p$) and coherent ($\kappa_c$) components are displayed in the top-right corner of each panel.

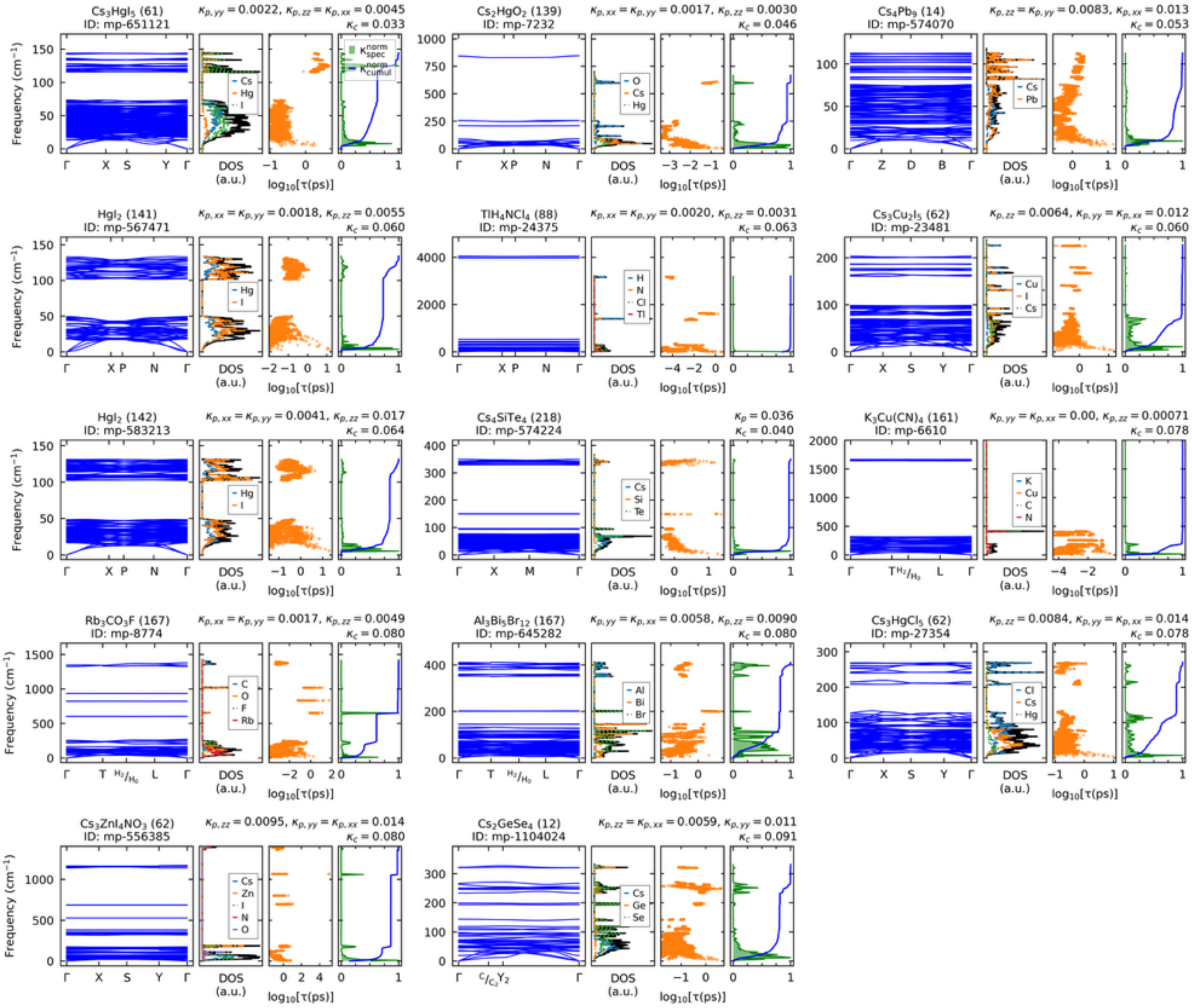

Fig. S3. List of phonon properties for materials with low thermal conductivity ($< 0.1\ \mathrm{Wm^{-1}K^{-1}}$), arranged in ascending order. See the caption of Fig. S2 for a detailed description of the figure.

## IV. COMPARISON BETWEEN PARTICLE AND COHERENT CONTRIBUTIONS TO HEAT TRANSPORT

Lattice thermal conductivity is governed by both particle-like phonon gas transport and wave-like tunneling mechanisms[2,14]. In the early stages of first-principles phonon analysis, calculating the coherent component arising from wave-like tunneling was challenging. However, since Simoncelli *et al*. proposed a unified theory (also known as the Wigner heat transport formula) to evaluate both the Peierls component ($\kappa_{\mathrm{p}}$) from particle-like phonons and the coherent component ($\kappa_{\mathrm{c}}$)[2], this approach has been widely adopted for analysis[15].

Among the materials calculated in Phonix, we identified that materials with a large coherent component ($\kappa_{\mathrm{c}}$) are polymorphs of SiC and related structures with the space group *P3m1* (156). While the relative contribution of $\kappa_{\mathrm{c}}$ to total heat transport remains small due to the relatively large $\kappa_{\mathrm{p}}$, the absolute magnitude of $\kappa_{\mathrm{c}}$ exceeds 50 Wm$^{-1}$K$^{-1}$. This significant coherent contribution arises from densely packed phonon branches, which result from the large number of atoms in the primitive cell. Their thermal conductivities ($\kappa_{\mathrm{p}}$ and $\kappa_{\mathrm{c}}$) and phonon dispersions are shown in Fig. S4.

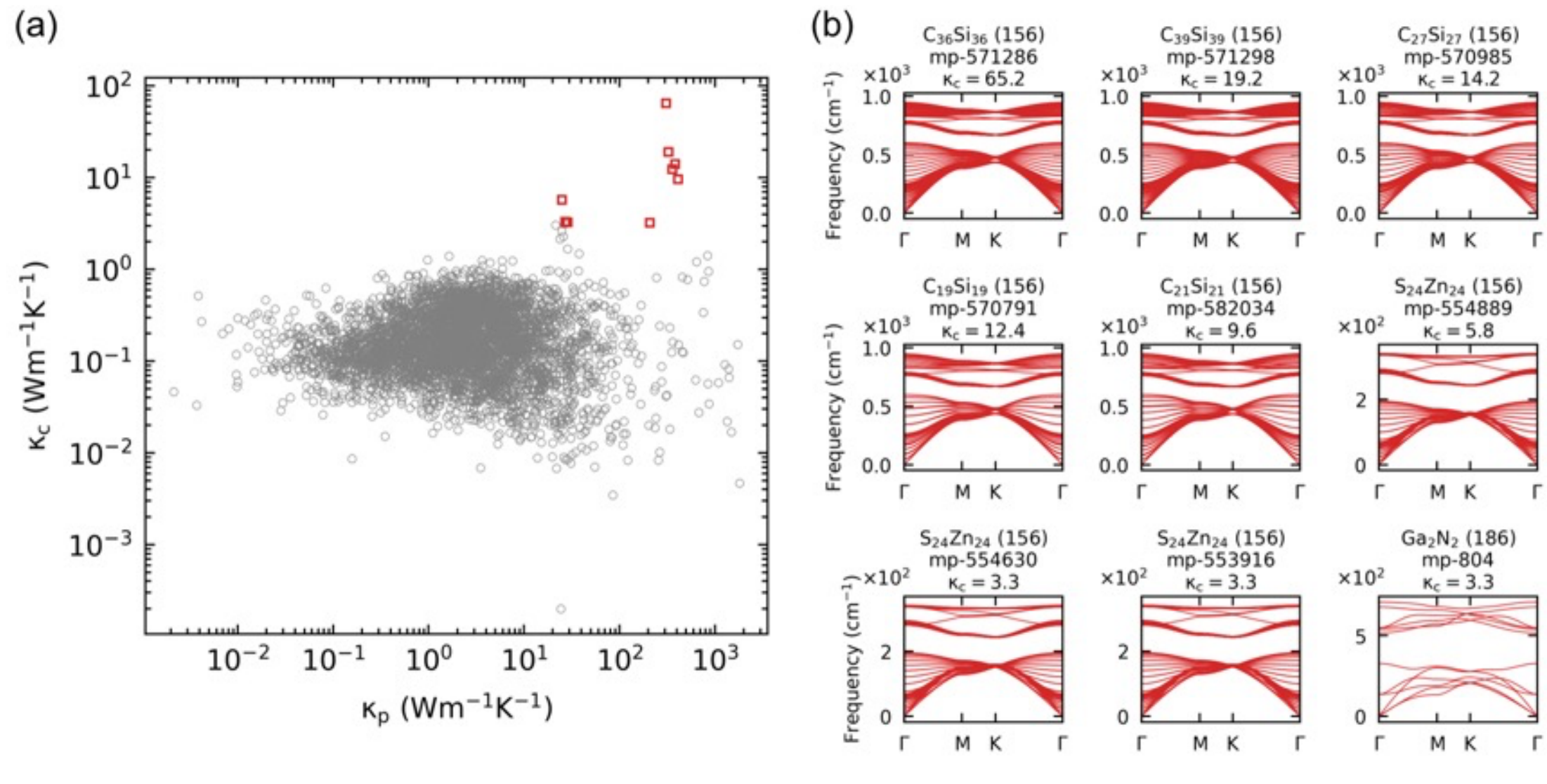

Fig. S4. Materials exhibiting large coherent thermal conductivity. (a) Distribution of Peierls ($\kappa_{\mathrm{p}}$) and coherent ($\kappa_{\mathrm{c}}$) thermal conductivity. Red markers indicate the top nine materials with the highest $\kappa_{\mathrm{c}}$. (b) Phonon dispersion for these nine materials.

## V. COMPARISON WITH PHONONDB

Phonon dispersions obtained in this study using automated calculations were compared with those in Phonondb[16]. Comparisons were possible for 155 out of 230 space groups. The Phonondb results were calculated using Phonopy[17] with displacement–force datasets stored in the database. The parameters used in this study for first-principles calculations, including the pseudopotentials, were largely the same as those in Phonondb, although the structures were re-relaxed in this study. The two sets of results are in excellent agreement, as shown in Fig. S5. The remaining discrepancies are likely attributable to differences in the relaxed structures, particularly the lattice constants. Overall, this comparison demonstrates the reliability of both datasets.

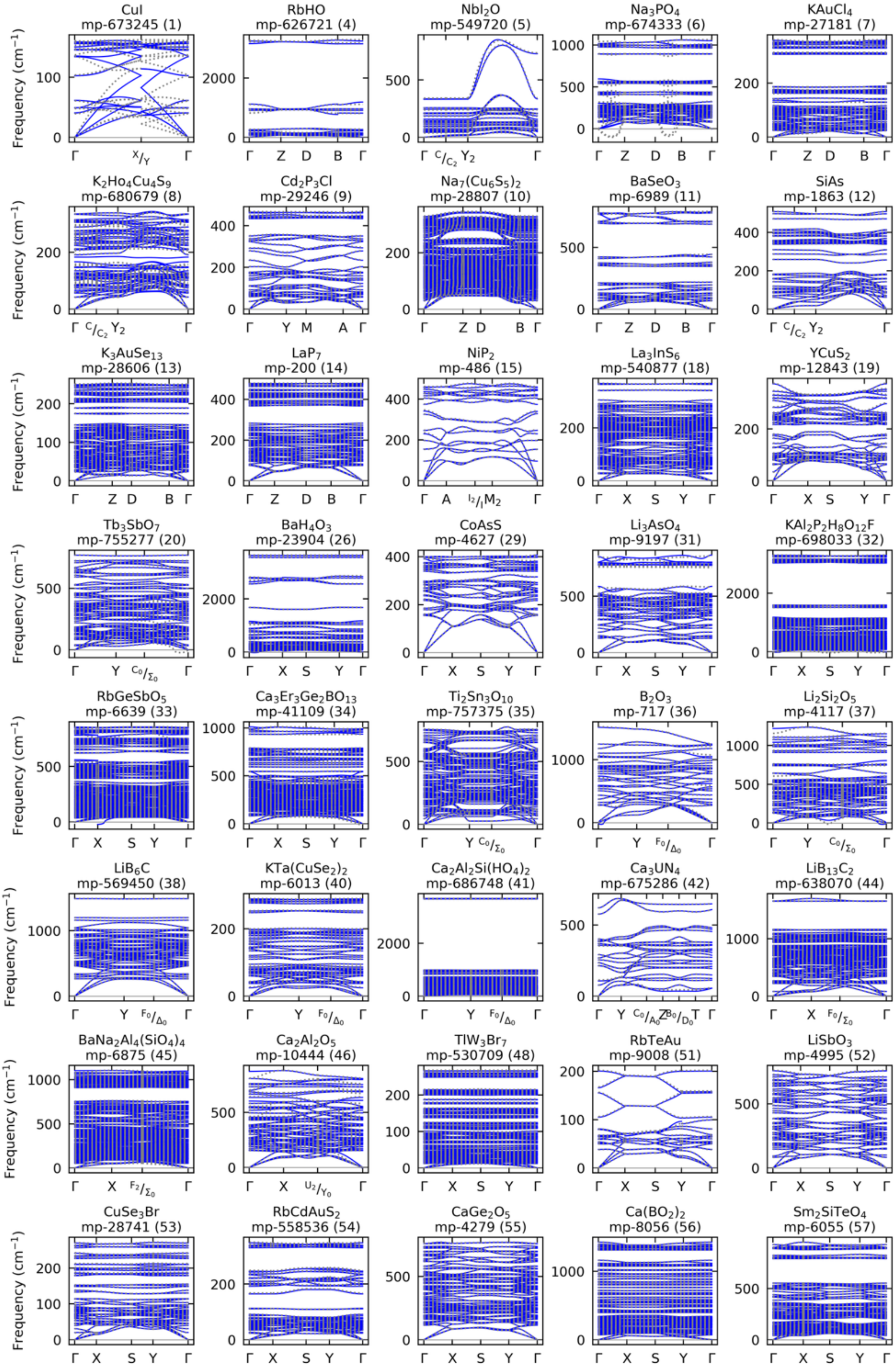
Frequency (cm$^{-1}$)
CuI mp-673245 (1)
RbHO mp-626721 (4)
NbI$_2$O mp-549720 (5)
Na$_3$PO$_4$ mp-674333 (6)
KAuCl$_4$ mp-27181 (7)
K$_2$Ho$_4$Cu$_4$S$_9$ mp-680679 (8)
Cd$_2$P$_3$Cl mp-29246 (9)
Na$_7$(Cu$_6$S$_5$)$_2$ mp-28807 (10)
BaSeO$_3$ mp-6989 (11)
SiAs mp-1863 (12)
K$_3$AuSe$_{13}$ mp-28606 (13)
LaP$_7$ mp-200 (14)
NiP$_2$ mp-486 (15)
La$_3$InS$_6$ mp-540877 (18)
YCuS$_2$ mp-12843 (19)
Tb$_3$SbO$_7$ mp-755277 (20)
BaH$_4$O$_3$ mp-23904 (26)
CoAsS mp-4627 (29)
Li$_3$AsO$_4$ mp-9197 (31)
KAl$_2$P$_2$H$_8$O$_{12}$F mp-698033 (32)
RbGeSbO$_5$ mp-6639 (33)
Ca$_3$Er$_3$Ge$_2$BO$_{13}$ mp-41109 (34)
Ti$_2$Sn$_3$O$_{10}$ mp-757375 (35)
B$_2$O$_3$ mp-717 (36)
Li$_2$Si$_2$O$_5$ mp-4117 (37)
LiB$_6$C mp-569450 (38)
KTa(CuSe$_2$)$_2$ mp-6013 (40)
Ca$_2$Al$_2$Si(HO$_4$)$_2$ mp-686748 (41)
Ca$_3$UN$_4$ mp-675286 (42)
LiB$_{13}$C$_2$ mp-638070 (44)
BaNa$_2$Al$_4$(SiO$_4$)$_4$ mp-6875 (45)
Ca$_2$Al$_2$O$_5$ mp-10444 (46)
TlW$_3$Br$_7$ mp-530709 (48)
RbTeAu mp-9008 (51)
LiSbO$_3$ mp-4995 (52)
CuSe$_3$Br mp-28741 (53)
RbCdAuS$_2$ mp-558536 (54)
CaGe$_2$O$_5$ mp-4279 (55)
Ca(BO$_2$)$_2$ mp-8056 (56)
Sm$_2$SiTeO$_4$ mp-6055 (57)

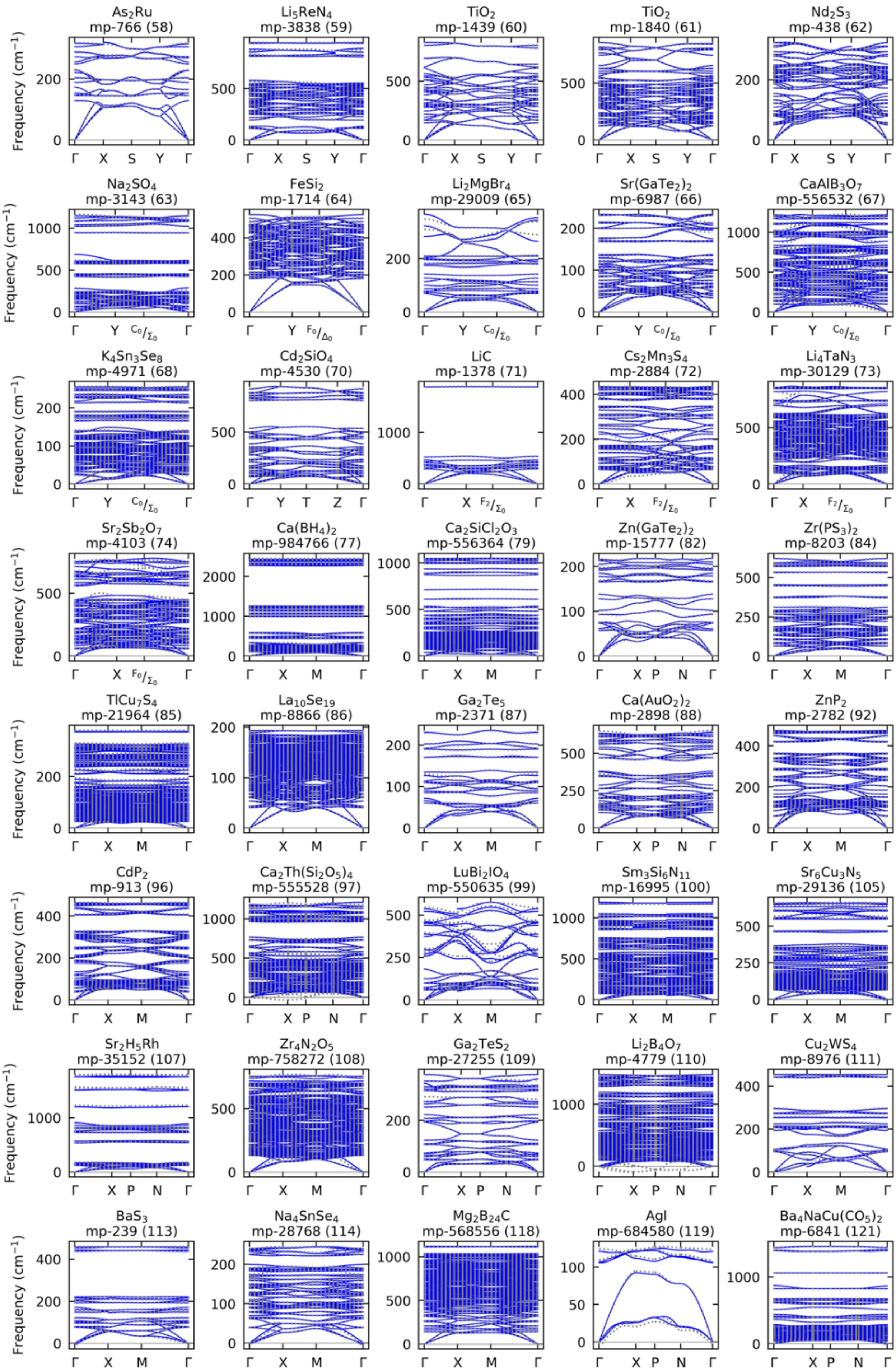
As2Ru mp-766 (58)
Li5ReN4 mp-3838 (59)
TiO2 mp-1439 (60)
TiO2 mp-1840 (61)
Nd2S3 mp-438 (62)
Na2SO4 mp-3143 (63)
FeSi2 mp-1714 (64)
Li2MgBr4 mp-29009 (65)
Sr(GaTe2)2 mp-6987 (66)
CaAlB3O7 mp-556532 (67)
K4Sn3Se8 mp-4971 (68)
Cd2SiO4 mp-4530 (70)
LiC mp-1378 (71)
Cs2Mn3S4 mp-2884 (72)
Li4TaN3 mp-30129 (73)
Sr2Sb2O7 mp-4103 (74)
Ca(BH4)2 mp-984766 (77)
Ca2SiCl2O3 mp-556364 (79)
Zn(GaTe2)2 mp-15777 (82)
Zr(PS3)2 mp-8203 (84)
TlCu7S4 mp-21964 (85)
La10Se19 mp-8866 (86)
Ga2Te5 mp-2371 (87)
Ca(AuO2)2 mp-2898 (88)
ZnP2 mp-2782 (92)
CdP2 mp-913 (96)
Ca2Th(Si2O5)4 mp-555528 (97)
LuBi2IO4 mp-550635 (99)
Sm3Si6N11 mp-16995 (100)
Sr6Cu3N5 mp-29136 (105)
Sr2H5Rh mp-35152 (107)
Zr4N2O5 mp-758272 (108)
Ga2TeS2 mp-27255 (109)
Li2B4O7 mp-4779 (110)
Cu2WS4 mp-8976 (111)
BaS3 mp-239 (113)
Na4SnSe4 mp-28768 (114)
Mg2B24C mp-568556 (118)
AgI mp-684580 (119)
Ba4NaCu(CO5)2 mp-6841 (121)
Frequency (cm−1)
Γ X S Y Γ
Γ Y C0/Σ0 Γ
Γ Y F0/Δ0 Γ
Γ Y T Z Γ
Γ X F2/Σ0 Γ
Γ X F0/Σ0 Γ
Γ X M Γ
Γ X P N Γ

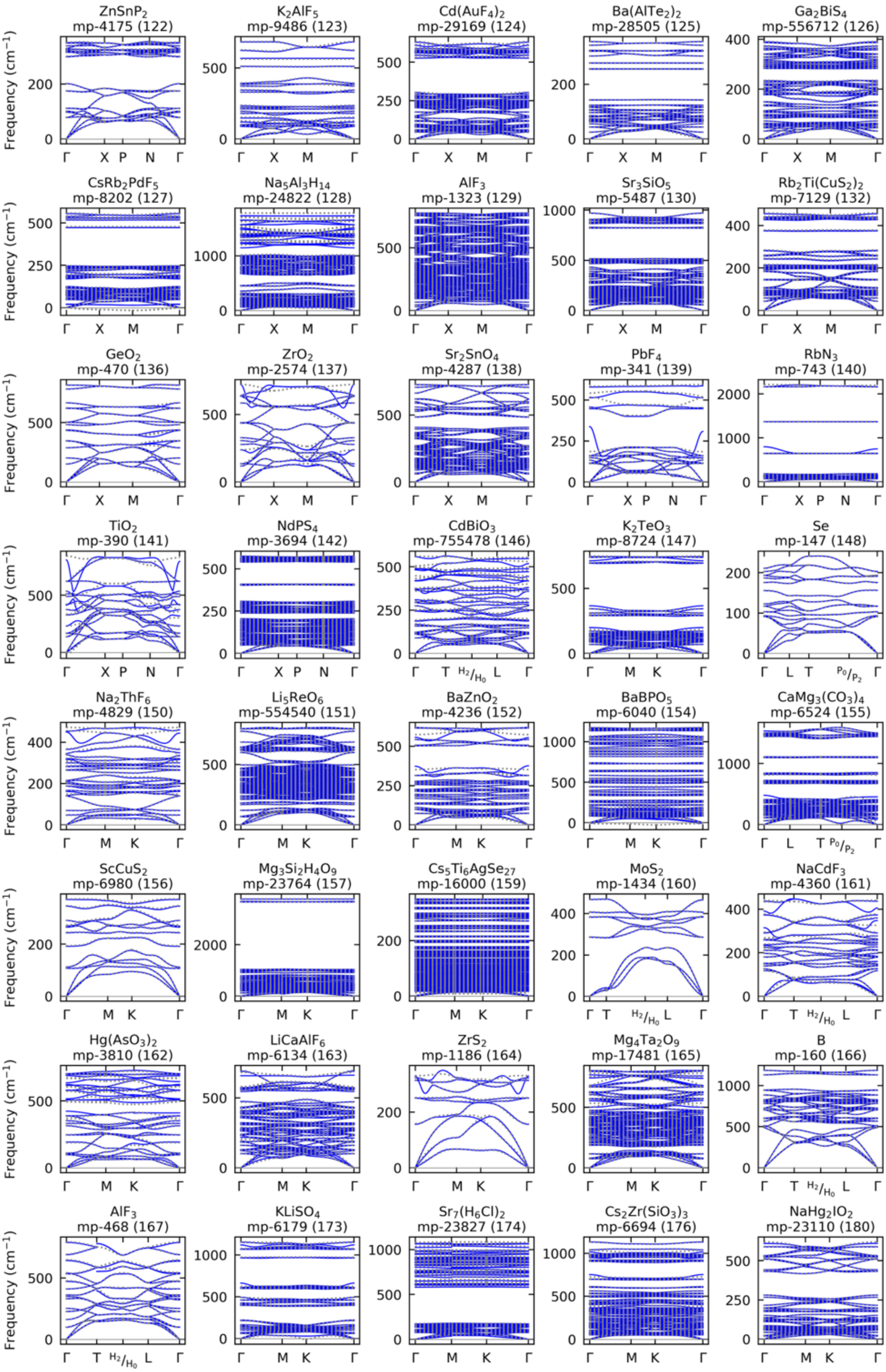
Frequency (cm$^{-1}$)
$ZnSnP_2$ mp-4175 (122)
$K_2AlF_5$ mp-9486 (123)
$Cd(AuF_4)_2$ mp-29169 (124)
$Ba(AlTe_2)_2$ mp-28505 (125)
$Ga_2BiS_4$ mp-556712 (126)
$CsRb_2PdF_5$ mp-8202 (127)
$Na_5Al_3H_{14}$ mp-24822 (128)
$AlF_3$ mp-1323 (129)
$Sr_3SiO_5$ mp-5487 (130)
$Rb_2Ti(CuS_2)_2$ mp-7129 (132)
$GeO_2$ mp-470 (136)
$ZrO_2$ mp-2574 (137)
$Sr_2SnO_4$ mp-4287 (138)
$PbF_4$ mp-341 (139)
$RbN_3$ mp-743 (140)
$TiO_2$ mp-390 (141)
$NdPS_4$ mp-3694 (142)
$CdBiO_3$ mp-755478 (146)
$K_2TeO_3$ mp-8724 (147)
Se mp-147 (148)
$Na_2ThF_6$ mp-4829 (150)
$Li_5ReO_6$ mp-554540 (151)
$BaZnO_2$ mp-4236 (152)
$BaBPO_5$ mp-6040 (154)
$CaMg_3(CO_3)_4$ mp-6524 (155)
$ScCuS_2$ mp-6980 (156)
$Mg_3Si_2H_4O_9$ mp-23764 (157)
$Cs_5Ti_6AgSe_{27}$ mp-16000 (159)
$MoS_2$ mp-1434 (160)
$NaCdF_3$ mp-4360 (161)
$Hg(AsO_3)_2$ mp-3810 (162)
$LiCaAlF_6$ mp-6134 (163)
$ZrS_2$ mp-1186 (164)
$Mg_4Ta_2O_9$ mp-17481 (165)
B mp-160 (166)
$AlF_3$ mp-468 (167)
$KLiSO_4$ mp-6179 (173)
$Sr_7(H_6Cl)_2$ mp-23827 (174)
$Cs_2Zr(SiO_3)_3$ mp-6694 (176)
$NaHg_2IO_2$ mp-23110 (180)

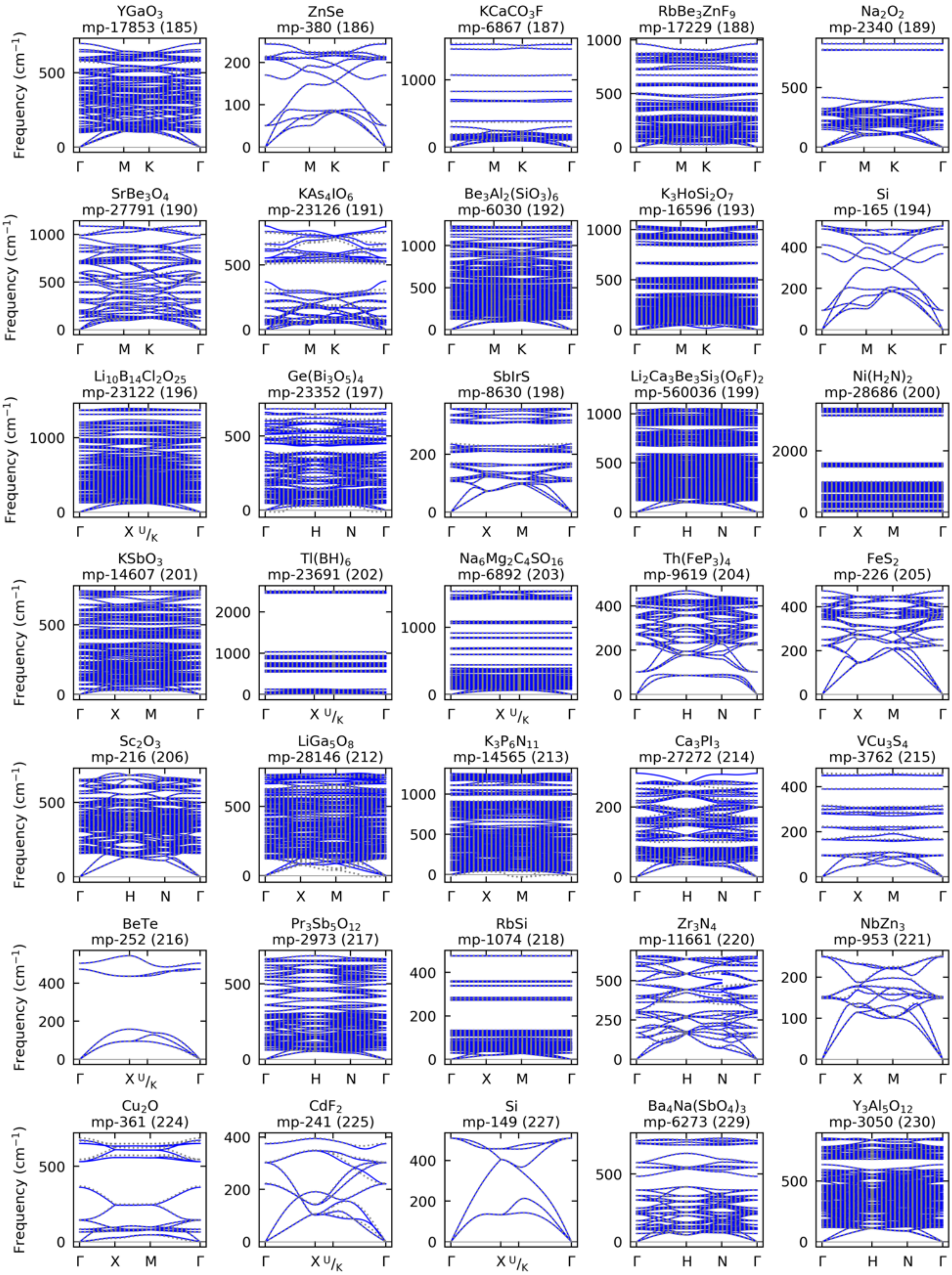

Fig. S5. Comparison of phonon dispersions obtained in this study (blue solid line) with those reported in Phonondb (grey dotted line) for all available space groups. The chemical formula, material ID in Materials Project, and space group number (in parentheses) are shown in the top of each panel.

## VI. COMPARISON WITH EXPERIMENTAL RESULTS

The calculated room-temperature lattice thermal conductivities ($\kappa_{\mathrm{lat}}^{\mathrm{RT}}$) were compared with experimental results for 103 single-crystal compounds, comprising 156 data points, as shown in Fig. S6. In this figure, the deviation of the calculated data ($\kappa_{\mathrm{DFT}}$) from the experimental data ($\kappa_{\mathrm{exp}}$) is evaluated in different formats, as described below.

Mean absolute error (MAE) in log scale:

$$\mathrm{MAE\ (log\,scale)} = \frac{1}{N}\sum\left(\log_{10}\kappa_{\mathrm{exp}} - \log_{10}\kappa_{\mathrm{DFT}}\right)$$

coefficient of determination ($R^2$):

$$R^2 = 1 - \frac{\sum\left(\kappa_{\mathrm{exp}} - \kappa_{\mathrm{DFT}}\right)^2}{\sum\left(\kappa_{\mathrm{exp}} - \overline{\kappa_{\mathrm{exp}}}\right)^2}$$

symmetry mean absolute percentage error (SMAPE):

$$\mathrm{SMAPE} = \frac{100}{N}\sum\frac{|\kappa_{\mathrm{DFP}} - \kappa_{\mathrm{exp}}|}{\left(|\kappa_{\mathrm{exp}}| + |\kappa_{\mathrm{DFP}}|\right)/2}$$

mean symmetry relative error (MSRE):

$$\mathrm{MSRE} = \frac{2}{N}\sum\frac{|\kappa_{\mathrm{DFT}} - \kappa_{\mathrm{exp}}|}{\kappa_{\mathrm{DFT}} + \kappa_{\mathrm{exp}}}$$

where *N* is the number of data points.

While certain materials—for example, those with significant long-range interactions such as PbX (X = S, Se, Te)—exhibit relatively large discrepancies between calculated and experimental values, the calculated data overall show good agreement with the experimental results. For materials with large discrepancies, including PbX and $ZnGeP_2$, the computational conditions employed in the automated calculations may not be sufficiently accurate; therefore, careful treatment with higher-level computational settings is required for more reliable predictions, as discussed in Supplementary Sec. VIII-A. Experimental data used in Fig. S6  are listed in Table S2.

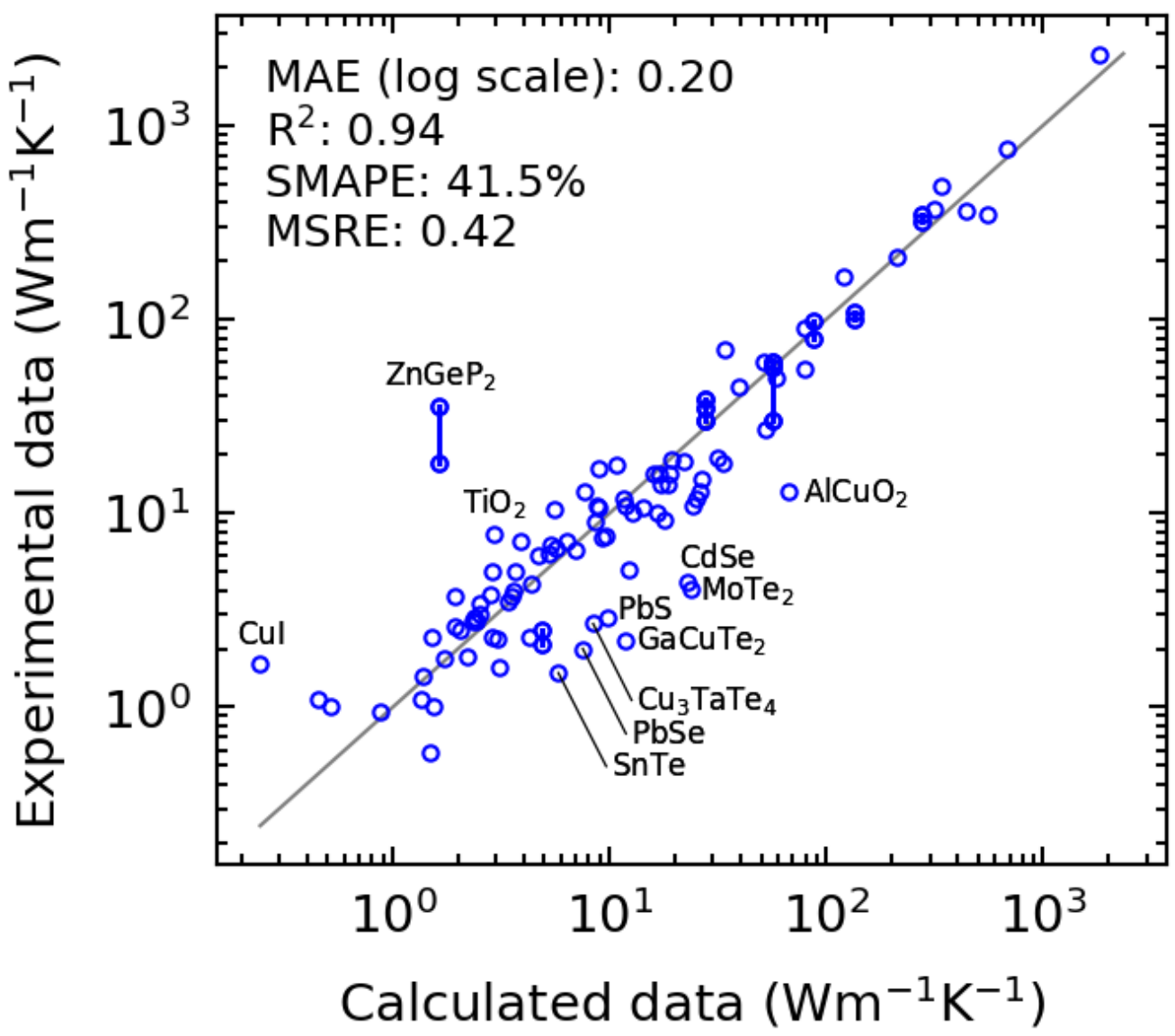

Fig. S6 Room-temperature lattice thermal conductivity of single-crystal compounds compared with experimental and computational data. The dataset includes 103 materials and 156 data points. Chemical formulas of materials exhibiting significant discrepancies are indicated in the figure.

Table S2 List of experimental data in Fig. S6.

| Material ID | Formula | Space-group | $\kappa_{lat}$ (W m$^{-1}$ K$^{-1}$) | Reference | DOI or book title |
|---|---|---|---|---|---|
| mp-66 | C | 227 | 2340 | Morelli (2005)[18] | DOI: 10.1007/0-387-25100-6_2 |
| mp-149 | Si | 227 | 166 | Morelli (2005)[18] | DOI: 10.1007/0-387-25100-6_2 |
| mp-216 | Sc2O3 | 206 | 17 | Li (2005)[19] | DOI:10.1111/j.1551-2916.2005.00162.x |
| mp-241 | CdF2 | 225 | 4.3 | Popov (2010)[20] | DOI: 10.1134/S1063783410030091 |
| mp-361 | Cu2O | 224 | 5 | Lide (2004)[21] | RC handbook of chemistry and physics |
| mp-406 | CdTe | 216 | 7.5 | Morelli (2005)[18] | DOI: 10.1007/0-387-25100-6_2 |
| mp-463 | KF | 225 | 6.43 | Morelli (2005)[18] | DOI: 10.1007/0-387-25100-6_2 |
| mp-602 | MoTe2 | 194 | 4.02 | Miller (2017) [22] | DOI: 10.1021/acs.chemmater.6b04179 |
| mp-643 | ThO2 | 225 | 14 | Mann (2010)[23] | DOI: 10.1021/cg901308f |
| mp-661 | AlN | 186 | 319 | Slack (1987)[24] | DOI: 10.1016/0022-3697(87)90153-3 |
| mp-661 | AlN | 186 | 350 | Morelli (2005)[18] | DOI: 10.1007/0-387-25100-6_2 |
| mp-672 | CdS | 186 | 16 | Morelli (2005)[18] | DOI: 10.1007/0-387-25100-6_2 |
| mp-682 | NaF | 225 | 18.4 | Morelli (2005)[18] | DOI: 10.1007/0-387-25100-6_2 |
| mp-804 | GaN | 186 | 210 | Morelli (2005)[18] | DOI: 10.1007/0-387-25100-6_2 |
| mp-856 | SnO2 | 136 | 98/55 | Turkes (1980)[25] | DOI: 10.1088/0022-3719/13/26/015. |
| mp-981 | SrF2 | 225 | 9.035 | Popov (2010)[20] | DOI: 10.1134/S1063783410030091 |
| mp-1029 | BaF2 | 225 | 6.8 | Popov (2010)[20] | DOI: 10.1134/S1063783410030091 |
| mp-1138 | LiF | 225 | 17.6 | Morelli (2005)[18] | DOI: 10.1007/0-387-25100-6_2 |
| mp-1143 | Al2O3 | 167 | 35 | Lide (2004)[21] | CRC handbook of chemistry and physics |
| mp-1143 | Al2O3 | 167 | 30 | Slack (1962)[26] | DOI: 10.1103/PhysRev.126.427 |
| mp-1143 | Al2O3 | 167 | 39 | Langenbert (2016)[27] | DOI: 10.1063/1.4966220 |
| mp-1190 | ZnSe | 216 | 19 | Morelli (2005)[18] | DOI: 10.1007/0-387-25100-6_2 |
| mp-1239 | IrSb3 | 204 | 16 | Slack (1994)[28] | DOI: 10.1063/1.357750 |
| mp-1265 | MgO | 225 | 60 | Morelli (2005)[18] | DOI: 10.1007/0-387-25100-6_2 |
| mp-1265 | MgO | 225 | 57 | Langenberg (2016)[27] | DOI: 10.1063/1.4966220 |
| mp-1265 | MgO | 225 | 30 | Slifka (1998)[29] | DOI: 10.6028/jres.103.021 |
| mp-1317 | CoSb3 | 204 | 10 | Toberer (2011)[30] | DOI: 10.1039/C1JM11754H |
| mp-1342 | BaO | 225 | 2.3 | Surplice (1963)[31] | DOI: 10.1088/0508-3443/14/10/333 |
| mp-1479 | BP | 216 | 350 | Morelli (2005)[18] | DOI: 10.1007/0-387-25100-6_2 |
| mp-1550 | AlP | 216 | 90 | Morelli (2005)[18] | DOI: 10.1007/0-387-25100-6_2 |
| mp-1639 | BN | 216 | 760 | Morelli (2005)[18] | DOI: 10.1007/0-387-25100-6_2 |
| mp-1883 | SnTe | 225 | 1.5 | Toberer (2011)[30] | DOI: 10.1039/C1JM11754H |
| mp-1960 | Li2O | 225 | 11 | Takahashi (1980)[32] | DOI: 10.1016/0022-3115(80)90036-7 |
| mp-1986 | ZnO | 216 | 50 | Morelli (2005)[18] | DOI: 10.1007/0-387-25100-6_2 |
| mp-2133 | ZnO | 186 | 60 | Morelli (2005)[18] | DOI: 10.1007/0-387-25100-6_2 |
| mp-2172 | AlAs | 216 | 80 | Steigmeier (1966)[33] | DOI: 10.1103/PhysRev.141.767 |
| mp-2172 | AlAs | 216 | 98 | Morelli (2005)[18] | DOI: 10.1007/0-387-25100-6_2 |
| mp-2176 | ZnTe | 216 | 18 | Morelli (2005)[18] | DOI: 10.1007/0-387-25100-6_2 |
| mp-2201 | PbSe | 225 | 2 | Morelli (2005)[18] | DOI: 10.1007/0-387-25100-6_2 |
| mp-2201 | PbSe | 225 | 1.6 | El-Sharkawy (1983)[13] | DOI: 10.1007/BF00502357 |
| mp-2469 | CdS | 216 | 16 | Morelli (2005)[18] | DOI: 10.1007/0-387-25100-6_2 |
| mp-2472 | SrO | 225 | 12 | Morelli (2005)[18] | DOI: 10.1007/0-387-25100-6_2 |
| mp-2490 | GaP | 216 | 110 | Morelli (2005)[18] | DOI: 10.1007/0-387-25100-6_2 |
| mp-2490 | GaP | 216 | 100 | Morelli (2005)[18] | DOI: 10.1007/0-387-25100-6_2 |
| mp-2534 | GaAs | 216 | 45 | Morelli (2005)[18] | DOI: 10.1007/0-387-25100-6_2 |
| mp-2542 | BeO | 186 | 370 | Morelli (2005)[18] | DOI: 10.1007/0-387-25100-6_2 |
| mp-2605 | CaO | 225 | 30 | Morelli (2005)[18] | DOI: 10.1007/0-387-25100-6_2 |
| mp-2624 | AlSb | 216 | 56 | Morelli (2005)[18] | DOI: 10.1007/0-387-25100-6_2 |
| mp-2652 | Y2O3 | 206 | 14 | Fornasiero et al., (2000)[34] | DOI: 10.1016/S0272-8842(99)00101-7 |
| mp-2657 | TiO2 | 136 | 7.8 | Langenbert (2016)[27] | DOI: 10.1063/1.4966220 |
| mp-2691 | CdSe | 216 | 4.4 | Toberer (2011)[30] | DOI: 10.1039/C1JM11754H |
| mp-2741 | CaF2 | 225 | 9.91 | Popov (2010)[20] | DOI: 10.1134/S1063783410030091 |
| mp-3050 | Y3Al5O12 | 230 | 10.4 | Langenberg (2016)[27] | DOI: 10.1063/1.4966220 |
| mp-3196 | NdGaO3 | 62 | 6 | Langenberg (2016)[27] | DOI: 10.1063/1.4966220 |
| mp-3359 | Ba2SnO4 | 139 | 6.13 | Miller (2017)[22] | DOI: 10.1021/acs.chemmater.6b04179 |
| mp-3536 | MgAl2O4 | 227 | 19.4 | Langenberg (1962)[27] | DOI: 10.1063/1.4966220 |
| mp-3595 | ZnSiAs2 | 122 | 14 | Beasley (1994)[35] | DOI: 10.1364/AO.33.001000 |
| mp-3668 | CdGeP2 | 122 | 11 | Beasley (1994)[35] | DOI: 10.1364/AO.33.001000 |
| mp-3748 | AlCuO2 | 166 | 13 | Yanagiya (2010)[36] | DOI: 10.3390/ma3010318 |
| mp-3792 | YAlO3 | 62 | 10.7 | Langenberg (2016)[27] | DOI: 10.1063/1.4966220 |
| mp-3839 | GaCuTe2 | 122 | 2.2 | Toberer (2011)[30] | DOI: 10.1039/C1JM11754H |
| mp-4524 | ZnGeP2 | 122 | 35.5 | Beasley (1994)[35] | DOI: 10.1364/AO.33.001000 |
| mp-4524 | ZnGeP2 | 122 | 18 | Masumoto (1966) [37] | DOI: 10.1016/0022-3697(66)90124-7 |
| mp-4840 | GaCuSe2 | 122 | 12.9 | Beasley (1994)[35] | DOI: 10.1364/AO.33.001000 |
| mp-5238 | GaCuS2 | 122 | 5.09 | Beasley (1994)[35] | DOI: 10.1364/AO.33.001000 |
| mp-5342 | AgGaS2 | 122 | 1.45 | Beasley (1994)[35] | DOI: 10.1364/AO.33.001000 |
| mp-5518 | GaAgSe2 | 122 | 0.26 | Garbato (1979([38] | DOI: 10.1016/0009-2614(79)80529-1 |
| mp-5837 | LaGaO3 | 62 | 5 | Langenberg (2016)[27] | DOI: 10.1063/1.4966220 |
| mp-5967 | TiCoSb | 216 | 14 | Sekimoto (2005)[39] | DOI: 10.2320/matertrans.46.1481 |
| mp-5967 | TiCoSb | 216 | 25 | Xia (2000)[40] | DOI: 10.1063/1.1305829 |
| mp-5967 | TiCoSb | 216 | 12 | Kawaharada (2004)[41] | DOI: 10.1016/j.jallcom.2004.04.096 |
| mp-7631 | SiC | 186 | 490 | Morelli (2005)[18] | DOI: 10.1007/0-387-25100-6_2 |
| mp-8017 | AlCuTe2 | 122 | 0.32 | Rincón (1995)[42] | DOI: 10.1002/pssa.2211470212 |
| mp-8062 | SiC | 216 | 360 | Harris (1995)[43] | Properties of Silicon Carbide |
| mp-9295 | Cu3TaTe4 | 215 | 2.69 | Miller (2017)[22] | DOI: 10.1021/acs.chemmater.6b04179 |
| mp-10695 | ZnS | 216 | 27 | Morelli (2005)[18] | DOI: 10.1007/0-387-25100-6_2 |
| mp-10760 | MgSe | 225 | 7.69 | Pawlak [44] | DOI: 10.1007/s10765-009-0590-6 |
| mp-11718 | RbF | 225 | 2.27 | Hakansson (1989)[45] | DOI: 10.1088/0953-8984/1/25/009 |
| mp-19717 | PbTe | 225 | 2.1 | Pei (2011)[46] | DOI: 10.1039/C0EE00456A |
| mp-19717 | PbTe | 225 | 2.5 | Morelli (2005)[18] | DOI: 10.1007/0-387-25100-6_2 |
| mp-19717 | PbTe | 225 | 2.0 | El-Sharkawy (1983)[13] | DOI: 10.1007/BF00502357 |
| mp-20194 | CeO2 | 225 | 10.8 | Jha (2019)[47] | DOI: 10.1201/b17045 |
| mp-20352 | Ba3In2O6 | 139 | 3.76 | Miller (2017)[22] | DOI: 10.1021/acs.chemmater.6b04179 |
| mp-20485 | InSe | 194 | 6.9 | El-Sharkawy (1983)[13] | DOI: 10.1016/0022-3697(70)90284-2 |
| mp-21276 | PbS | 225 | 2.9 | Morelli (2005)[18] | DOI: 10.1007/0-387-25100-6_2 |
| mp-21276 | PbS | 225 | 2.5 | El-Sharkawy (1983)[13] | DOI: 10.1007/BF00502357 |
| mp-22220 | LaLuO3 | 62 | 2.5 | Langenberg (2016)[27] | DOI: 10.1063/1.4966220 |
| mp-22862 | NaCl | 225 | 7.1 | Morelli (2005)[18] | DOI: 10.1007/0-387-25100-6_2 |
| mp-22867 | RbBr | 225 | 3.8 | Morelli (2005)[18] | DOI: 10.1007/0-387-25100-6_2 |
| mp-22895 | CuI | 225 | 1.68 | Lide (2004)[21] | CRC handbook of chemistry and physics |
| mp-22898 | KI | 225 | 2.6 | Morelli (2005)[18] | DOI: 10.1007/0-387-25100-6_2 |
| mp-22903 | RbI | 225 | 2.3 | Morelli (2005)[18] | DOI: 10.1007/0-387-25100-6_2 |
| mp-22916 | NaBr | 225 | 2.8 | Morelli (2005)[18] | DOI: 10.1007/0-387-25100-6_2 |
| mp-22922 | AgCl | 225 | 1 | Beasley (1994)[35] | DOI: 10.1364/AO.33.001000 |
| mp-23193 | KCl | 225 | 7.1 | Morelli (2005)[18] | DOI: 10.1007/0-387-25100-6_2 |
| mp-23209 | SrCl2 | 225 | 2.3 | Moore (1985)[48] | DOI: 10.1007/978-1-4684-4916-7_12 |
| mp-23231 | AgBr | 225 | 1.1 | Mamran (2007)[49] | DOI: 10.1088/0022-3727/40/3/027 |
| mp-23251 | KBr | 225 | 3.4 | Morelli (2005)[18] | DOI: 10.1007/0-387-25100-6_2 |
| mp-23259 | LiBr | 225 | 1.83 | Hakansson (1989)[45] | DOI: 10.1088/0953-8984/1/25/009. |
| mp-23268 | NaI | 225 | 1.8 | Morelli (2005)[18] | DOI: 10.1007/0-387-25100-6_2 |
| mp-23295 | RbCl | 225 | 2.8 | Morelli (2005)[18] | DOI: 10.1007/0-387-25100-6_2 |
| mp-23703 | LiH | 225 | 15 | Morelli (2005)[18] | DOI: 10.1007/0-387-25100-6_2 |
| mp-31117 | NdScO3 | 62 | 4 | Langenbert et al.[27] | DOI: 10.1063/1.4966220 |
| mp-31118 | SmScO3 | 62 | 3.5 | Langenbert et al.[27] | DOI: 10.1063/1.4966220 |
| mp-31119 | TbScO3 | 62 | 3 | Langenberg et al.[27] | DOI: 10.1063/1.4966220 |
| mp-31120 | DyScO3 | 62 | 2.9 | Langenberg et al.[27] | DOI: 10.1063/1.4966220 |
| mp-34202 | Bi2Te3 | 166 | 1.6 | Toberer (2011)[30] | DOI: 10.1039/C1JM11754H |
| mp-540688 | SrIn2O4 | 62 | 6.53 | Miller (2017)[22] | DOI: 10.1021/acs.chemmater.6b04179 |
| mp-559756 | PrScO3 | 62 | 3.7 | Langenberg (2016)[27] | DOI: 10.1063/1.4966220 |
| mp-567636 | VFeSb | 216 | 13 | Young (2000)[50] | DOI: 10.1063/1.371863 |
| mp-568560 | TlBr | 63 | 0.58 | Zhu (2021)[51] | DOI: 10.1039/D1EE00442E |
| mp-571222 | CsBr | 225 | 0.94 | Gerlich (2000)[52] | DOI: 10.1088/0022-3719/15/25/013 |
| mp-573697 | CsCl | 225 | 1 | Gerlich (2000)[52] | DOI: 10.1088/0022-3719/15/25/013 |
| mp-614603 | CsI | 225 | 1.1 | Gerlich (2000)[52] | DOI: 10.1088/0022-3719/15/25/013 |
| mp-755574 | LuAlO3 | 62 | 11 | Langenberg (2016)[27] | DOI: 10.1063/1.4966220 |
| mp-924130 | TiNiSn | 216 | 9.3 | Hohl (1999)[53] | DOI: 10.1088/0953-8984/11/7/004 |
| mp-1017439 | SrSiO3 | 221 | 10.1 | Xiao (2013)[54] | DOI: 10.1103/PhysRevB.96.161201 |

## VII. PRECAUTIONS FOR AUTOMATED CALCULATIONS

In high-throughput calculations, there is always a trade-off between computational accuracy and speed. Based on our knowledge and experience with first-principles phonon calculations, we have aimed to ensure the highest possible computational accuracy in the automated calculations conducted in this study. However, we found that some materials still exhibit implausible data, as shown in Fig. S7.

Implausible data led to excessively high $\kappa_{\mathrm{p}}$, exceeding a few thousand $\mathrm{Wm^{-1}K^{-1}}$. In most cases, this overestimation arises from excessively long phonon lifetimes in isolated flat branches or low-frequency acoustic modes, as well as from excessively high group velocities. To eliminate the excessively long lifetimes, the increase in **q**-mesh density and consideration of four-phonon interaction[55] are necessary. The computational cost of four-phonon interactions is excessively high, but machine-learning prediction[56] may be an effective strategy for accurately predicting high $\kappa_{\mathrm{p}}$ values. In contrast, excessively large group velocities are likely due to insufficient structural optimization. For example, implausible changes in phonon branches can be observed around the T point of $Sr_2CuBrO_2$ in Fig. S7. While these implausible data can be automatically handled in future versions of the software, they were manually excluded from the discussion in this study.

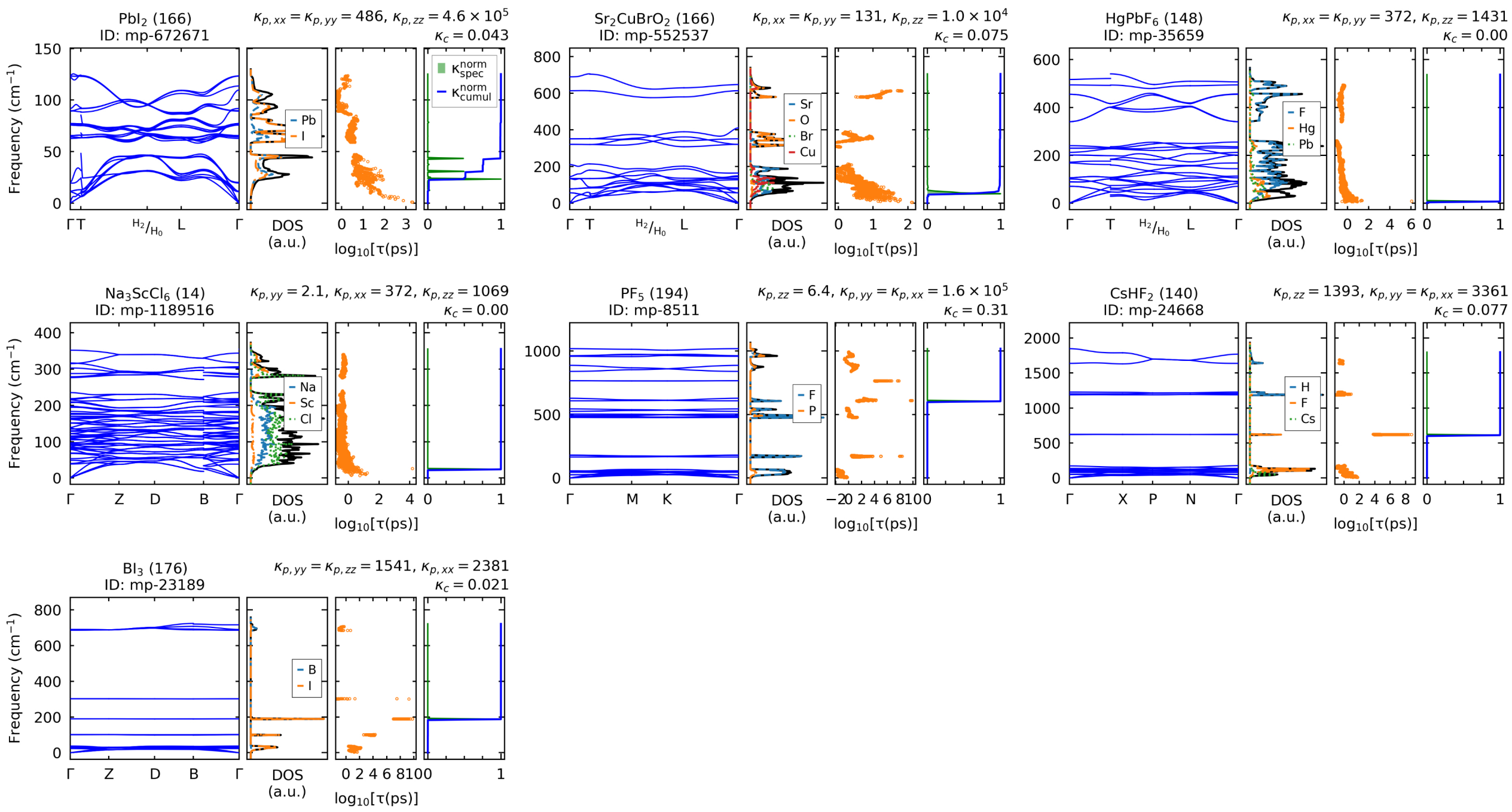

Fig. S7. Results of the automated calculations, including implausible data.

## VIII. COMPUTATIONAL ACCURACY

### A. Effect of Supercell Size on Phonon Properties

In the first loop of the automated calculations conducted in this study, the maximum number of atoms in a supercell was set to 150, and the supercell size for harmonic force constants (FCs) was subsequently increased if imaginary frequencies appeared. Each supercell was generated from the conventional cell so as to maximize the number of atoms within this limit. Because all possible harmonic FCs in the generated supercell were included in the phonon analysis, the maximum atomic distance of the harmonic FCs depended on the supercell size. In contrast, the cutoff distance for cubic FCs was set to the larger of 4.3 Å or the third-shortest atomic distance. The criterion for determining the cubic-FC cutoff was optimized based on insights gained during the automated calculations, ensuring that the dominant cubic FCs were captured for most materials. In this Supplementary section, we present calculations of phonon properties for selected materials using different supercell sizes. We demonstrate that the above criteria—namely, the choice of supercell size (which determines the harmonic-FC cutoff) and the cubic-FC cutoff length—were reasonable for most materials. Nevertheless, we note that these parameters may require careful verification when analyzing specific materials

We analyzed the size dependence of phonon properties in a set of randomly selected materials. Most of these materials showed no significant size dependence at the supercell size used in this study (≈150 atoms), as illustrated in Fig. S 8 (a). Specifically, the phonon dispersion, phonon lifetime, and temperature-dependent Peierls lattice thermal conductivity exhibited no appreciable size dependence. To further investigate size effects, we next focused on rocksalt PbX (X = S, Se, Te), which are well known to exhibit long-range interactions[57–60]. Figure S8(b) shows that the phonon dispersion does not converge with increasing supercell size in these compounds. While the phonon dispersion of PbTe exhibits a relatively moderate size dependence compared with PbS and PbSe, its thermal conductivity fluctuates considerably with changes in supercell size. Conversely, even for relatively small supercells ($2 \times 2 \times 2$ and $3 \times 3 \times 3$ supercells containing 64 and 216 atoms, respectively), the anharmonic phonon properties (phonon lifetime and thermal conductivity) of PbS and PbSe show no appreciable size dependence. These results demonstrate that the criteria adopted for cubic FCs ensure convergence in most materials, including those with significant long-range interactions, owing to the much shorter interaction range of cubic FCs compared with harmonic FCs[60]. In contrast, for harmonic FCs, convergence was not achieved even in large systems (e.g., a 5×5×5 supercell with 1,000 atoms). Therefore, the use of the density functional perturbation theory (DFPT) approach may be necessary to further improve computational accuracy.

Finally, examples of materials whose phonon properties did not converge within our criteria are shown in Fig. S8(c). In certain materials, slight changes in the phonon dispersion—typically in modes with large Grüneisen parameters—affect the resulting lattice thermal conductivity, as in the case of CsAu. In PrRh, the phonon dispersion exhibits an anomalous change with increasing supercell size, leading to the emergence of imaginary frequencies. For $ZnGeP_2$, the thermal conductivity fluctuates by an order of magnitude, despite meeting the criteria adopted in this study.

In summary, the computational criteria for the automated calculations were determined by balancing computational cost and accuracy. While most materials showed good convergence under different computational conditions, including supercell size, the dataset for certain materials contains substantial errors. Therefore, when using data for specific materials, it is necessary to examine the detailed results and, if required, to recalculate with more stringent conditions. It may also be of interest to use the Phonix database to optimize the computational conditions for each material.

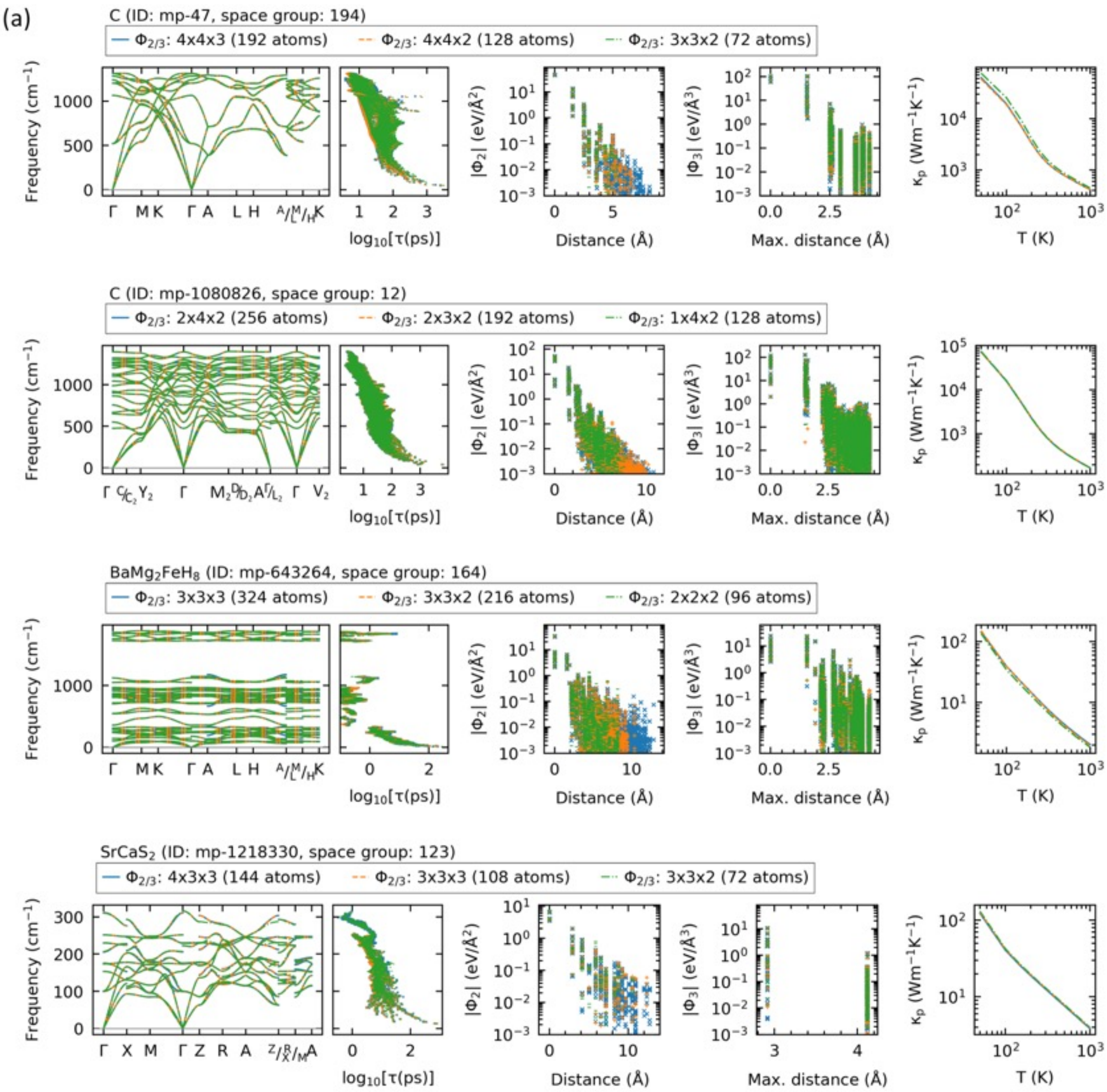

Fig. S8. Effect of supercell size on phonon properties. Each panel shows phonon dispersion, phonon lifetime on a logarithmic scale ($|\log_{10} \tau \text{ (ps)}|$), harmonic ($\Phi_2$) and cubic ($\Phi_3$) force constants, and Peierls lattice thermal conductivity ($\kappa_p$). At the top of each panel, the chemical formula, material ID, and space group number are shown. In the legend, each color denotes the supercell size (relative to the primitive cell) used for force constant calculations, as well as the number of atoms in the supercell. In several cases, different supercell sizes were used to compute $\Phi_2$ and $\Phi_3$ constants. (a) Examples of materials showing convergence of phonon properties with respect to the supercell size.

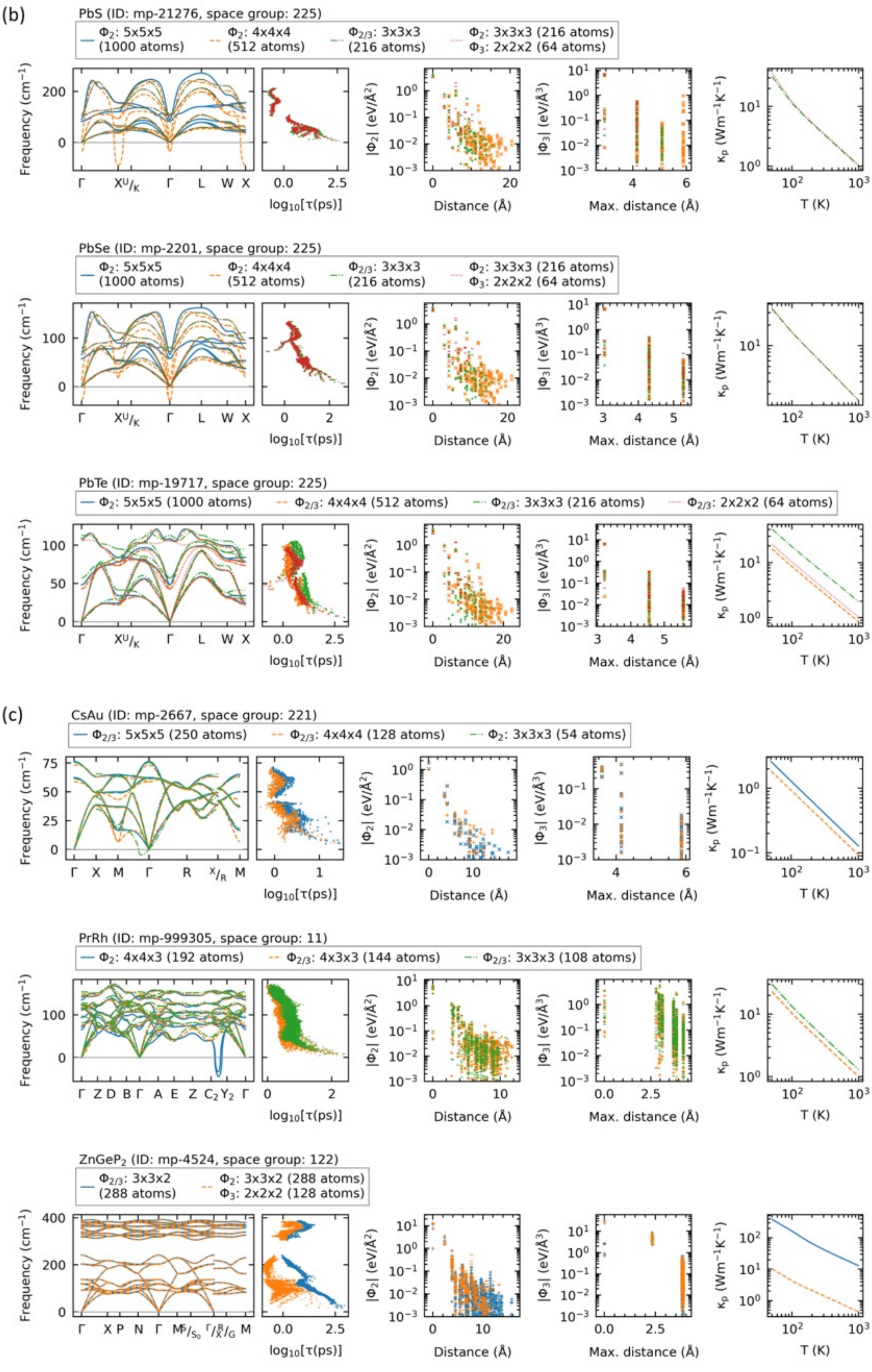

Fig. S8. (contd.) (b) Phonon properties of PbX (X = S, Se, Te) calculated for different supercell sizes. These compounds are known to exhibit significant long-term interactions. (c) Examples of materials with a significant size effect in phonon properties.

### B. Comparison between the Finite-Displacement Method and LASSO Regression

In the automated calculations performed in this study, the least absolute shrinkage and selection operator (LASSO) regression was applied when the number of suggested displacement patterns for the finite-displacement (FD) method exceeded 100; otherwise, the FD method was used. To obtain cubic force constants, one or two atoms were displaced in the FD method, whereas all atoms in the supercell were randomly displaced in the LASSO approach. The displacement magnitudes were set to 0.03 Å in both cases.

We compared the force constants and the resulting temperature-dependent thermal conductivity obtained from these approaches, as illustrated in Fig. S9. To reduce the computational cost, particularly for the FD method, the comparison was conducted for relatively simple materials. The force constants obtained with both approaches were in excellent agreement. Although certain materials may exhibit small discrepancies between the two approaches—for example, NaF (mp-682) in Fig. S9—the results are generally comparable. This comparison suggests that, although both methods were employed in this study, LASSO regression may be a suitable choice for future high-throughput calculations, as it reduces computational time and improves methodological consistency across materials.

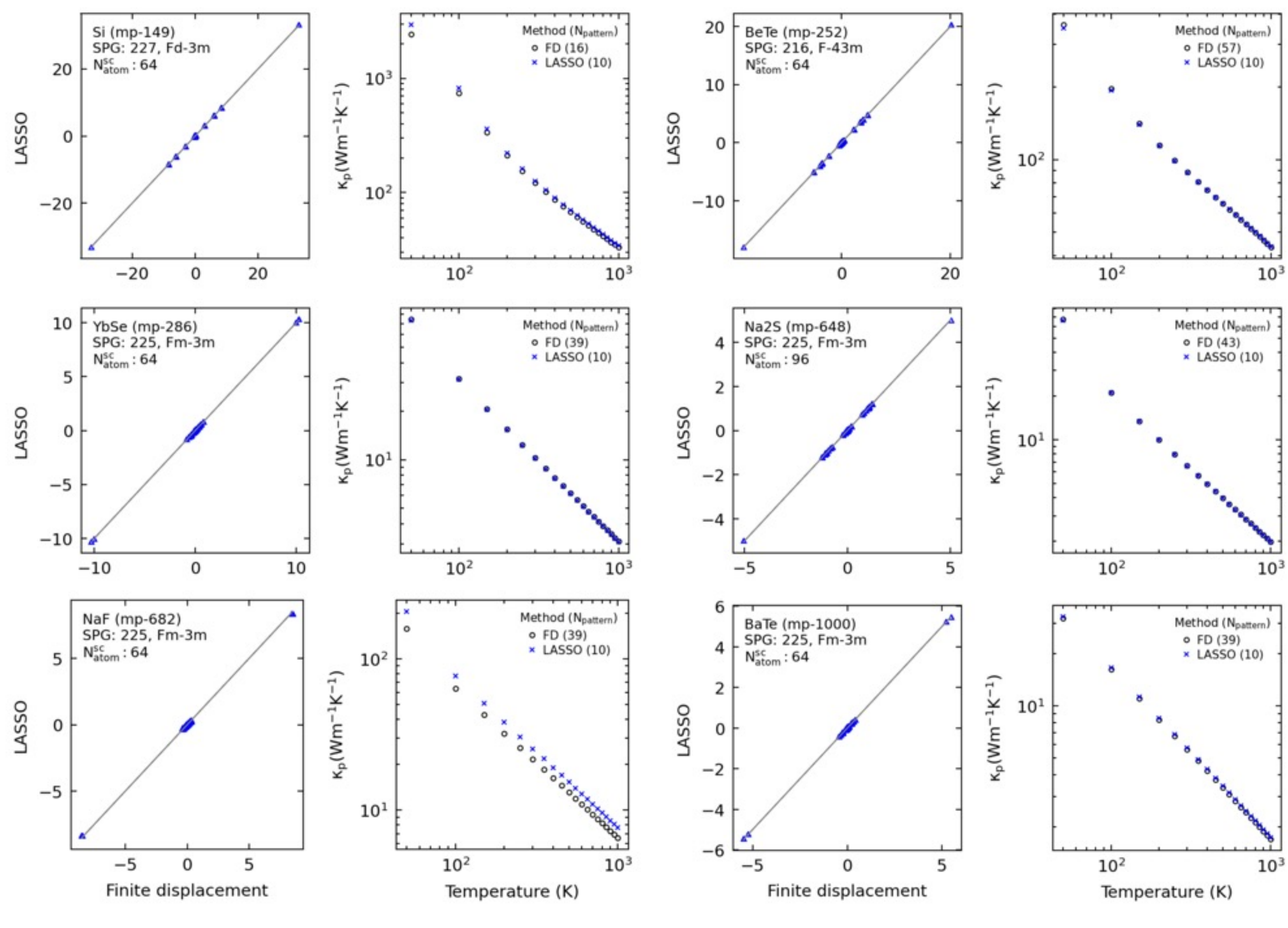

Fig. S9. Comparison of anharmonic phonon properties—force constants (left) and Peierls lattice thermal conductivity (right)—obtained using the finite displacement (FD) method and LASSO regression. In each left panel, the chemical formula, material ID (in parentheses), space group (SPG) number and name, and the number of atoms in the supercell are shown. In each right panel, the numbers of displacement patterns used for each method are indicated in parentheses.

## MACHINE LEARNING-BASED PREDICTION OF PHONON PROPERTIES

Using the Phonix database, we predicted the Peierls lattice thermal conductivity ($\kappa_{\mathrm{p}}$) and its cumulative value with respect to the mean free path ($\kappa_{\mathrm{cumul}}$), as mentioned in the main text. In addition to these properties, we also conducted the prediction for the coherent component ($\kappa_{\mathrm{c}}$), maximum phonon frequency ($\omega_{\mathrm{max}}$), and spectral $\kappa_{\mathrm{p}}$ as a function of frequency ($\kappa_{\mathrm{spec}}(\omega)$). For training of these properties, we used $\log_{10}(\kappa_{\mathrm{c}}+\varepsilon)$ (where $\varepsilon = 0.001\,\mathrm{Wm^{-1}K^{-1}}$), $\log_{10}\omega_{\mathrm{max}}$, and $\kappa_{\mathrm{spec}}(\omega)$ normalized by its maximum value, respectively. For $\kappa_{\mathrm{spec}}^{\mathrm{norm}}(\omega)$, the data were prepared from 0 to $\omega_{\mathrm{max}}$ with 51 points. In the study on phonon density of states (DOS) prediction[61], which we referred to for the prediction method using the Euclidean neural networks (e3nn)[62], DOS data were discretized into 51 points from 0 to 1,000 $\mathrm{cm^{-1}}$. In contrast, in our prediction, the frequency was normalized by its maximum value because the maximum frequency does not reach 1,000 $\mathrm{cm^{-1}}$ in most materials. Spectral thermal conductivity as a function of the absolute frequency can be obtained by combining the predicted values of $\omega_{\mathrm{max}}$ and $\kappa_{\mathrm{spec}}^{\mathrm{norm}}(\omega/\omega_{\mathrm{max}})$. For the machine learning technique, CGCNN[63] was applied to the prediction of $\kappa_{\mathrm{c}}$ and $\omega_{\mathrm{max}}$ and e3nn[62] was applied to $\kappa_{\mathrm{spec}}$.

We observed deep learning scaling laws for these three phonon properties, similar to those for $\kappa_{\mathrm{p}}$ and $\kappa_{\mathrm{cumul}}^{\mathrm{norm}}(\Lambda)$, as discussed in the main text. In every case, the prediction accuracy increased as the training dataset size increased. The scaling factor was 0.077 to 0.30, which are compatible magnitude as those observed for large language model (0.095)[64] and deep learning for materials in a previous study (0.21)[65]. Notably, since the prediction accuracy for $\omega_{\mathrm{max}}$ was remarkably high (with an MAE of 0.03 in log-scaled $\omega_{\mathrm{max}}$ ($\mathrm{cm^{-1}}$)), integrating the predictions of $\omega_{\mathrm{max}}$ and $\kappa_{\mathrm{spec}}^{\mathrm{norm}}(\omega/\omega_{\mathrm{max}})$ may provide more valuable information than directly predicting $\kappa_{\mathrm{spec}}^{\mathrm{norm}}(\omega)$, which approaches zero at high frequencies.

While the prediction accuracy for $\kappa_{\mathrm{spec}}(\omega)$ (with an MAE of 0.070) was slightly lower than that for $\kappa_{\mathrm{cumul}}(\Lambda)$ (with an MAE of 0.064), Phonix enables accurate prediction of the spectral anharmonic property, as shown in Fig. S10. The figure illustrates that spectral thermal conductivity was excellently predicted for half of the materials (green and blue), whereas predicting spectra with multiple peaks remains challenging, as indicated by the orange and red lines in Fig. S10.

Consequently, these data clearly show that Phonix opens up multiple possibilities for predicting not only phonon properties but also their interactions with other quasiparticles or related properties. At the same time, further increasing the dataset size or developing advanced machine learning techniques for phonon properties remains essential.

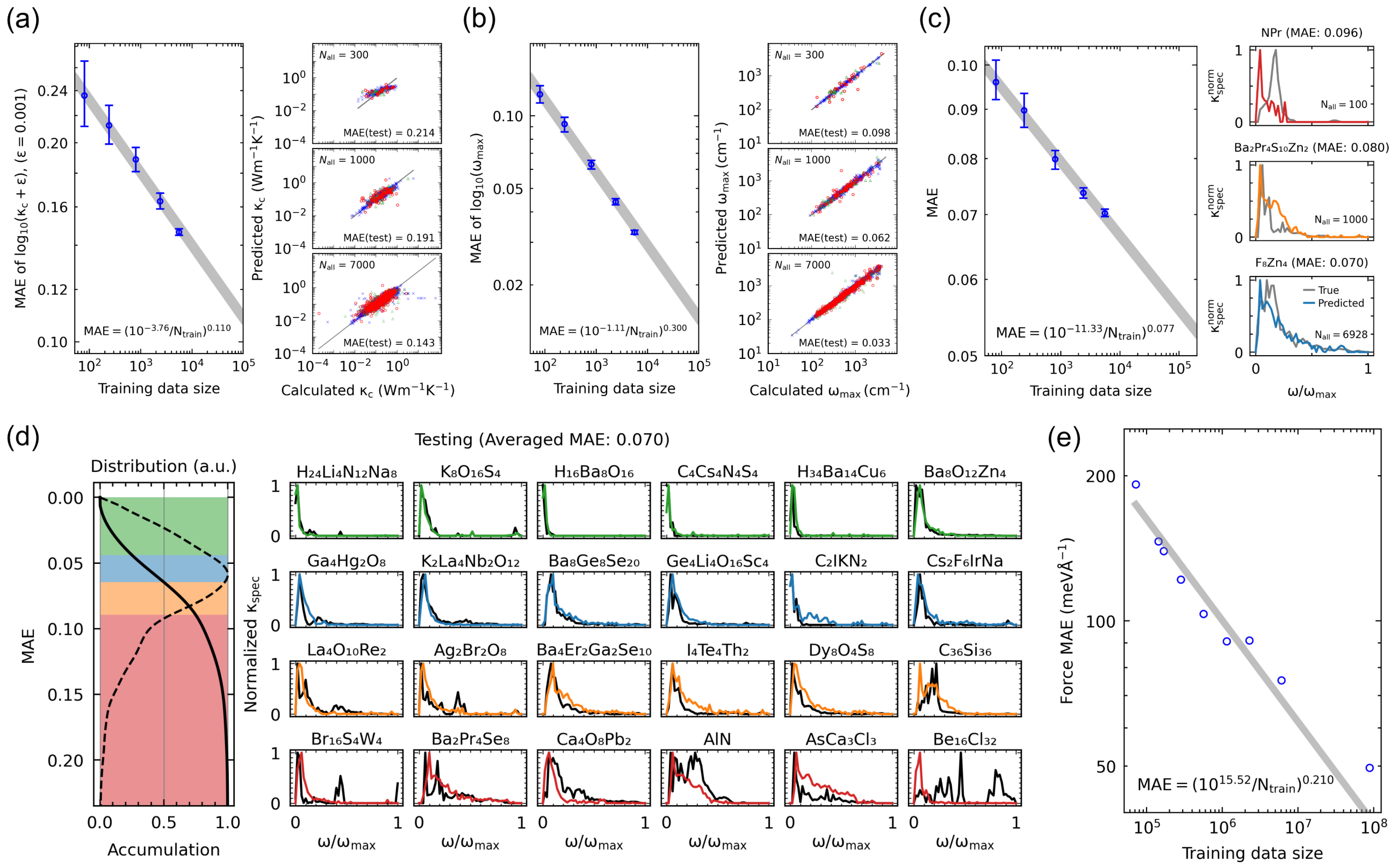

Fig. S10. Deep learning scaling law for phonon properties with respect to the training data size. (a) Coherent thermal conductivity ($\kappa_c$), (b) maximum frequency ($\omega_{max}$), and (c) normalized spectral Peierls thermal conductivity ($\kappa_{spec}^{norm}$) were predicted. The MAEs were evaluated with $\log_{10}(\kappa_c + \varepsilon)$, where $\varepsilon = 0.001$ Wm$^{-1}$K$^{-1}$, $\log_{10}[\omega_{max}\,(\mathrm{cm}^{-1})]$], and $\kappa_{spec}^{norm}$, respectively. (d) $\kappa_{spec}^{norm}$ prediction for the test data using all available data ($N_{all} \approx 7{,}000$). See the caption of Fig. 3 in the main text for details. (e) Prediction accuracy for atomic forces as a function of training data size, reproduced from Ref.[65] to compare scalability.

## IX. LIST OF SCREENED MATERIALS WITH HIGH AND LOW THERMAL CONDUCTIVITY

Using the prediction models built with Phonix, we screened materials with high and low $\kappa_{\mathrm{p}}$ from the GNoME database, as discussed in the main text. Through this screening, we identified three materials with $\kappa_{\mathrm{lat}}^{\mathrm{3ph}} > 200\ \mathrm{Wm^{-1}K^{-1}}$ and nine materials with $\kappa_{\mathrm{lat}}^{\mathrm{3ph}} < 0.2\ \mathrm{Wm^{-1}K^{-1}}$. Their phonon properties are presented in Fig. S11. For the three high-κ materials, we considered four-phonon scattering in addition to three-phonon scattering using the ALAMODE package. Although four-phonon scattering significantly reduces the thermal conductivity—for example, from 277 to 78 $\mathrm{Wm^{-1}K^{-1}}$ in NpPH—the three compounds still exhibit relatively high values (>50 $\mathrm{Wm^{-1}K^{-1}}$) compared with the entire set of materials in Phonix (see Fig. 2(b)).

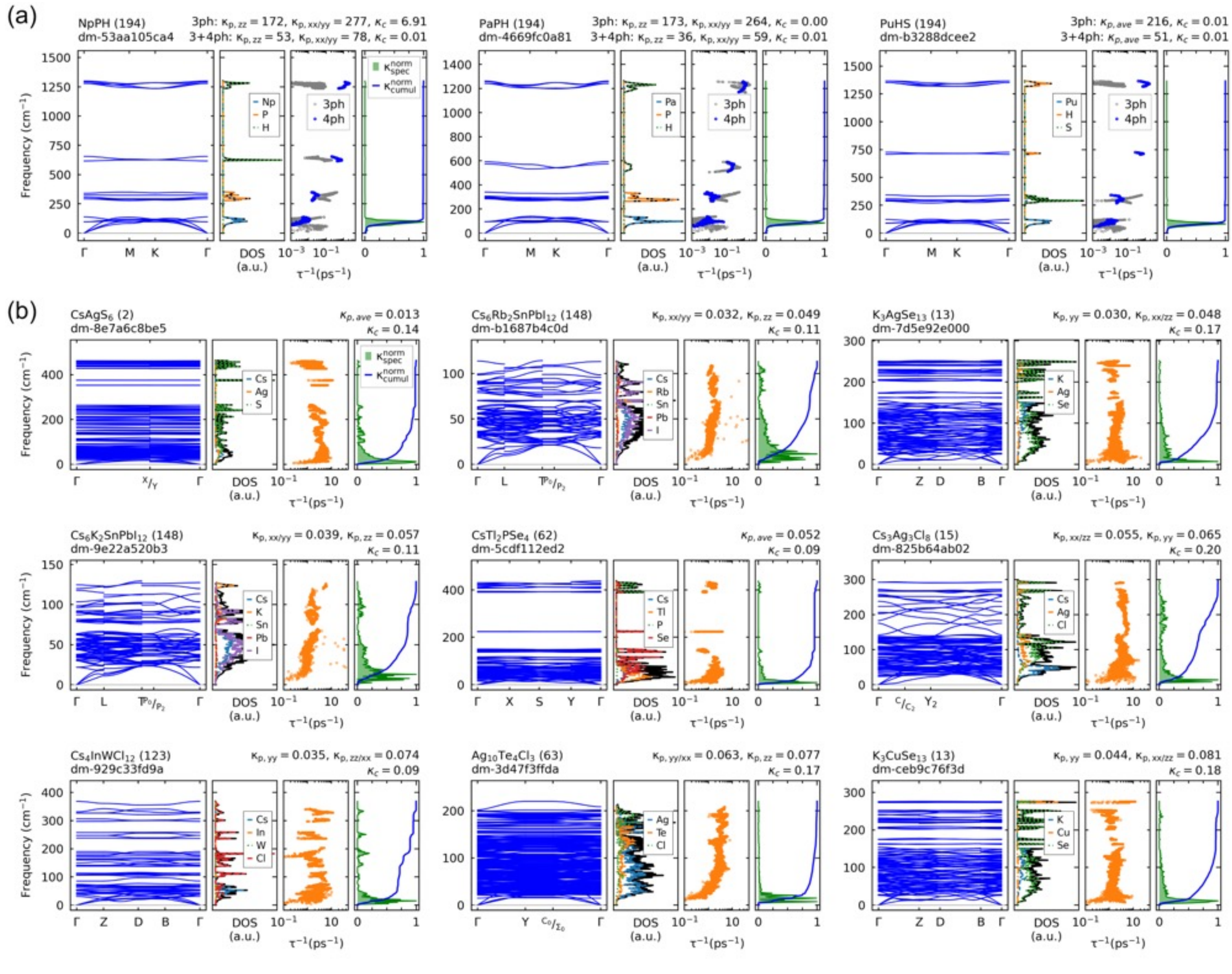

Fig. S11. Novel materials exhibiting extreme thermal conductivity. (a) A list of materials with $\kappa_{\mathrm{lat}} > 200\ \mathrm{Wm^{-1}K^{-1}}$ and (b) materials with $\kappa_{\mathrm{lat}} < 0.2\ \mathrm{Wm^{-1}K^{-1}}$. The materials were identified through screening of the GNoME database[65].